%% file: nuGMSB_Inf_3.tex
\documentclass[prd,superscriptaddress,twocolumn,nofootinbib,showpacs]{revtex4-2}

\usepackage{amsfonts,amssymb,amsmath}
\usepackage{color}
\usepackage{graphicx}
\usepackage{dcolumn}
\usepackage{bm}

\newcommand{\C}[1]{{\mathcal #1}}

\newcommand{\BS}[1]{{\boldsymbol #1}}

\newcommand{\Tr}{\mathop{\rm Tr}}

\newcommand{\half}{\frac 12}

\newcommand{\Slash}[1]{{\ooalign{\hfil#1\hfil\crcr\raise.167ex\hbox{/}}}}

\begin{document}




\title{Inflation and type III seesaw mechanism in
$\nu$-gauge mediated supersymmetry breaking}

\author{Shinsuke Kawai}
\affiliation{Department of Physics, 
Sungkyunkwan University,
Suwon 16419, Republic of Korea}
\author{Nobuchika Okada}
\affiliation{
Department of Physics and Astronomy, 
University of Alabama, 
Tuscaloosa, AL35487, USA} 
\date{\today}


\begin{abstract}
We discuss realization of cosmic inflation in the $\nu$-gauge mediated supersymmetry breaking scenario, in which a set of ${\BS 2\BS 4}$-dimensional chiral superfields 
responsible for the type III seesaw mechanism play the role of the messenger fields in gauge mediation.
Using the data from neutrino oscillations, we show that the model satisfies constraints from the lepton flavor violation, perturbativity of the unified gauge couplings, the observed abundance of dark matter as well as the Higgs mass of 125.1 GeV.
The predicted spectrum of the cosmic microwave background radiation fits well with the observation. 
We also comment on the falsifiability of this scenario by future experiments.
\end{abstract}


\maketitle


\section{Introduction}\label{sec:Intro}

Neutrino oscillations suggest that there is new physics beyond the Standard Model.
A natural extension of the Standard Model to account for neutrino oscillations is to include a set of new fields that generate the neutrino masses, via the so-called seesaw mechanism \cite{Minkowski:1977sc,Yanagida:1979as,GellMann:1980vs,Mohapatra:1979ia}.
Supersymmetry improves the Standard Model in several respects.
It controls the radiative corrections of the scalar sector while realizing electroweak symmetry breaking.
It also provides a natural candidate of dark matter, and predicts natural unification of the forces at the grand unification scale.
Since supersymmetry is observed to be broken at low energy, understanding of the breaking mechanism is essential for phenomenological studies.
Gauge mediated supersymmetry breaking (GMSB) is an elegant proposal for supersymmetry breaking, in which the effects of supersymmetry breaking are transmitted from the hidden sector to the visible sector by means of messenger fields that are charged under the Standard Model gauge group.

The seesaw mechanism and the GMSB are usually treated in different contexts, but it would be certainly appealing if the two phenomena are explained as arising from a common origin.
The seesaw fields (the right-handed neutrinos in the case of the type I seesaw) and the messenger fields in the GMSB are both assumed to be heavy fields that are integrated out to give low energy effective theories; 
then it is not particularly odd to speculate that they may have the same origin. 
There have been proposals to implement the GMSB and the seesaw mechanism in a unified framework \cite{Joaquim:2006uz,Joaquim:2006mn,Mohapatra:2007js,Mohapatra:2008gz,Mohapatra:2008wx};
these are called the $\nu$-GMSB models.
The model of $\nu$-GMSB proposed of Ref.~\cite{Mohapatra:2008wx} uses a set of ${\BS 2\BS 4}$-dimensional fields added to the minimal supersymmetric $SU(5)$ grand unified theory (GUT), and incorporates the type III seesaw mechanism.
The ${\BS 2\BS 4}$-dimensional fields, which we call $\Sigma_i$ in this paper, play a pivotal role in the construction of the model.
The type III seesaw mechanism is realized by triplet fields with zero hypercharge, that play the same role as the right-handed neutrinos in the type I seesaw mechanism. 
These triplet fields are contained in $\Sigma_i$.
For the successful type III seesaw mechanism, at least two triplet fields, and thus at least two ${\BS 2\BS 4}$-dimensional fields $(\Sigma_1,\Sigma_2)$ are needed.
We mainly focus on this {\em minimal seesaw} case in this paper, since three or more generations of $\Sigma_i$ will not be compatible with the perturbativity and the lepton flavor violation constraints, as will be discussed in Sec.~\ref{sec:perturbativity} and Sec.~\ref{sec:LFV}.
The $\Sigma_i$ fields also play the role of the messenger fields of GMSB.
Since these are full $SU(5)$ multiplets, they do not spoil the gauge coupling unification.
The dual role of $\Sigma_i$, as the seesaw fields and the messenger fields, is the key feature of the $\nu$-GMSB model.

The purpose of this paper is twofold. 
The model of the $\nu$-GMSB \cite{Mohapatra:2008wx} appeared in 2008.
While the basic proposal of this model is still valid, there has been an enormous progress in experimental physics since then, including the discovery of the Higgs boson and the precision measurements of the neutrino oscillation data.
Clearly, reinvestigation of the model in the light of the recent data is necessary.
The second aim of the paper is to investigate possible inflationary scenarios within this particle physics model.
We shall consider two possible directions of the inflationary trajectory, one along the messenger direction and the other in the $LH_u$ direction.
We discuss phenomenological consistency in each case, and show that the inflationary scenario in the $LH_u$ direction is a viable option, while the one in the messenger direction is not.

The rest of the paper is organized as follows.
The next section reviews the $\nu$-GMSB model with the type III seesaw mechanism \cite{Mohapatra:2008wx}.
In Sec.~\ref{sec:perturbativity} and Sec.~\ref{sec:LFV} the constraints from the perturbativity of the gauge couplings and the lepton flavor violation are discussed. 
In Sec.~\ref{sec:inflation} we investigate two possible scenarios of inflation, as well as their viability.
We conclude in Sec.~\ref{sec:final} with comments.
The formulae of the neutrino mass matrix are collected in the Appendix.

\section{The $\nu$-GMSB scenario}\label{sec:nuGMSB}

The $\nu$-GMSB model with the type III seesaw mechanism \cite{Mohapatra:2008wx} is realized by a set of ${\BS 2\BS 4}$-dimensional chiral superfields $\Sigma_i$ added to the minimal supersymmetric $SU(5)$ model.
The latter consists of two sets of matter fields 
$\overline{F_i}$ (${\BS 5}^*$ of $SU(5)$), $T_i$ (${\BS 1\BS 0}$) 
and three types of Higgs fields $\overline F_H$ (${\BS 5}^*$),
$F_H$ (${\BS 5}$) and $\Phi$ (${\BS 2\BS 4}$).
The index $i=1,2,3$ is for the generations of the matter fields, but we also consider the restricted case $i=1,2$ for $\Sigma_i$ in the minimal seesaw model.
Denoting the hidden sector\footnote{
We neglect the dynamics of the hidden sector which can be nontrivial in general \cite{Arai:2010ds,Arai:2010qe}.
} collectively by an $SU(5)$ singlet chiral superfield $S$ (${\BS 1}$), the superpotential of the $\nu$-GMSB model reads
\begin{align}\label{eqn:Wfull}
  W_{\nu\rm GMSB} &=W_{SU(5)}+Y_D^{ij}\overline F_j\Sigma_i F_H+ yS\Tr(\Sigma_i\Sigma_i),
\end{align}
where
\begin{align}\label{eqn:WSU5}
  W_{SU(5)} &=Y_d^{ij}\overline F_i T_j\overline F_H+Y_u^{ij}T_iT_jF_H
  +M_H\overline F_H F_H\crcr
  &+\lambda_1\overline F_H\Phi F_H+M_\Phi\Tr(\Phi^2)+\lambda_2\Tr(\Phi^3)
\end{align}
is the superpotential of the $SU(5)$ part.
We are concerned with the physics of supersymmetry breaking, the seesaw mechanism and the implementation of cosmic inflation, for which 
$T_i$, $\overline F_H$ and $\Phi$ are not important;
it is then sufficient to consider the superpotential 
\begin{align}\label{eqn:W}
  W= \sqrt\frac 52\, Y_D^{ij}\overline{F_j}\Sigma_i F_H+(yS+M)\Tr \left(\Sigma_i\Sigma_i\right)
\end{align}
that includes $\overline F_i$ (${\BS 5}^*$), $F_H$ (${\BS 5}$), $\Sigma_i$ (${\BS 2\BS 4}$) and $S$ (${\BS 1}$).
The factor of $\sqrt{5/2}$ in the first term has been introduced for later convenience.
In \eqref{eqn:Wfull}, supersymmetry is assumed to be broken by nonvanishing 
hidden sector $S$ field and its F-term,
\begin{align}
  \langle S\rangle=y^{-1}M\neq 0,\quad
  \langle F_S\rangle \neq 0.
\end{align}
In \eqref{eqn:W}, $S$ has been redefined by shifting $yS\to yS+M$ so that the supersymmetry breaking vacuum is now at $\langle S\rangle =0$.
The parameter $M$ is the mass of the fermionic components of $\Sigma_i$.
The squared masses of the scalar components $\Sigma_i^\pm\equiv (\Sigma_i\pm\Sigma_i^\dagger)/\sqrt 2$ are
$M_{\Sigma^\pm}^2=M^2\pm F_S$.
The deviation of the bosonic masses from the fermionic mass signals the breaking of supersymmetry.
Since $\Sigma_i$ are charged under the $SU(5)$ gauge group, the effect of supersymmetry breaking is communicated to the visible sector.
Thus the fields $\Sigma_i$ are the {\em messengers} of GMSB.

The $SU(5)$ adjoint $\Sigma_i$ are decomposed under the Standard Model gauge group $SU(3)_c\otimes SU(2)_L\otimes U(1)_Y$ as
\begin{align}\label{eqn:MMdecomp}
  {\BS 2\BS 4}=({\BS 1},{\BS 1},0) 
  \oplus ({\BS 8},{\BS 1},0) 
  \oplus ({\BS 1},{\BS 3},0) \crcr
  \oplus ({\BS 3},{\BS 2},-\frac 56) 
  \oplus ({\BS 3}^*,{\BS 2},+\frac 56).
\end{align}
Denoting the singlet $({\BS 1},{\BS 1},0)$ as $\widehat S_i$
and the triplet $({\BS 1},{\BS 3},0)$ as $\widehat T_i$,
one may write
\begin{align}
  \Sigma_i\supset\frac{\widehat S_i}{\sqrt{60}}{\rm diag}(2,2,2,-3,-3)
  +\frac{\widehat T_i}{2}{\rm diag}(0,0,0,1,-1).
\end{align}
The superpotential is then written
\begin{align}
  W\supset 
  &\sqrt\frac 52 Y_D^{ij}(e_j,\; -\nu_j)\widehat S_i\frac{-3}{\sqrt{60}}
  \left(\begin{array}{cc}
    1 & 0\\ 0 & 1
  \end{array}\right)
  \left(\begin{array}{c}
    H_u^+\\ H_u^0
  \end{array}\right)\crcr
  &+\sqrt\frac 52 Y_D^{ij}(e_j,\; -\nu_j)\widehat T_i\frac{1}{2}
  \left(\begin{array}{cc}
    1 & 0\\ 0 & -1
  \end{array}\right)
  \left(\begin{array}{c}
    H_u^+\\ H_u^0
  \end{array}\right)\crcr
  &+\half (yS+M)(\widehat S_i\widehat S_i+\widehat T_i\widehat T_i)\crcr
  \supset
  &\frac{3}{\sqrt{60}}\sqrt\frac 52 Y_D^{ij}\nu_j H_u^0\widehat S_i
  +\half\sqrt\frac 52 Y_D^{ij}\nu_j H_u^0\widehat T_i \crcr
  &+\half M (\widehat S_i\widehat S_i+\widehat T_i\widehat T_i).
\end{align}
%
%
In our conventions the generators of gauge groups are normalized as $\Tr(T^a\, T^b)=\half\delta^{ab}$.
Using the stationarity conditions
\begin{align}
  \frac{\partial W}{\partial\widehat S_i}
  =&\frac{3}{\sqrt{60}}\sqrt\frac 52 Y_D^{ij}\nu_j H_u^0
  +M\widehat S_i=0,\crcr
  \frac{\partial W}{\partial\widehat T_i}
  =&\half\sqrt\frac 52 Y_D^{ij}\nu_j H_u^0
  +M\widehat T_i=0,
\end{align}
to integrate out $\widehat S_i$ and $\widehat T_i$, one finds the effective superpotential
\begin{align}
  W_{\rm eff} =&
   -\half\frac{(Y_D^TY_D)^{ij}}{M}\nu_i\nu_jH_u^0H_u^0\crcr
  &=-\half\frac{(m_D^Tm_D)^{ij}}{M}\nu_i\nu_j,
\end{align}
where
\begin{align}\label{eqn:Weff}
  \langle H_u^0\rangle=\frac{v_u}{\sqrt 2},\qquad
  (m_D)^{ij}=Y_D^{ij}\cdot\frac{v_u}{\sqrt 2},
\end{align}
and $v_u=v\sin\beta$, $v = 246$ GeV.
One sees from this effective superpotential that the neutrino masses are given by the type III seesaw formula
\begin{align}\label{eqn:seesaw}
  (m_\nu)^{ij}=-\frac{v_u^2}{M}\left(Y_D^TY_D\right)^{ij}.
\end{align}

\begin{table}
\caption{\label{tab:table1}
Mass spectrum of the type III seesaw $\nu$-GMSB model in the units of GeV.
We used {\sc softsusy} 4.1.10 \cite{Allanach:2001kg}, with the values of the three parameters $N_5$, $M$, $\tan\beta$ as input and the remaining minimal GMSB parameter $\Lambda$ fixed by the condition that the Higgs mass is $m_h = 125.1$ GeV.
We chose $\mu>0$.
}
\begin{ruledtabular}
\begin{tabular}{c|c|c|c}
$\tan\beta$ & \multicolumn{3}{c}{10} \\
$N_5$ & \multicolumn{3}{c}{10} \\
\hline
$M$ & $10^{11}$ & $10^{12}$ & $10^{13}$ \\
$\Lambda$ & $1.801\times 10^5$ & $1.674\times 10^5$ & $1.569\times 10^5$ \\
\hline
$h_0$ & \multicolumn{3}{c}{125.1} \\
\hline
$H_0$ & 5417 & 5361 & 5305 \\
$A_0$ & 5417 & 5361 & 5305 \\
$H^\pm$ & 5418 & 5361 & 5306 \\
\hline
$\widetilde g$ & $1.057\times 10^4$ & 9894 & 9329 \\
$\widetilde\chi_{1,2}^0$ & 2479, 4361 & 2301, 4189 & 2154, 3937 \\
$\widetilde\chi_{3.4}^0$ & 4382, 4549 & 4350, 4372 & 4308, 4318 \\
$\widetilde\chi_{1,2}^\pm$ & 4361, 4549 & 4189, 4372 & 3937, 4318 \\
\hline
$\widetilde u$, $\widetilde c_{1,2}$ & 8771, 9223 & 8357, 8807 & 8001, 8450 \\
$\widetilde t_{1,2}$ & 7611, 8680 & 7159, 8249 & 6769, 7879 \\
$\widetilde d$, $\widetilde s_{1,2}$ & 8724, 9223 & 8306, 8807 & 7943, 8450 \\
$\widetilde b_{1,2}$ & 8671, 8701 & 8242, 8280 & 7872, 7916 \\
\hline
$\widetilde\nu_{e,\mu}$ & 3250 & 3193 & 3156 \\
$\widetilde\nu_\tau$ & 3245 & 3187 & 3149 \\
$\widetilde e$, $\widetilde\mu_{1,2}$ & 1746, 3252 & 1768, 3195 & 1809, 3157\\
$\widetilde\tau_{1,2}$ & 1725, 3246 & 1745, 3188 & 1785, 3150 \\
\end{tabular} 
\end{ruledtabular}
\end{table}

\section{Constraints from the perturbativity of the gauge couplings}\label{sec:perturbativity}

The seesaw mechanism requires at least two ${\BS 2\BS 4}$ dimensional fields $(\Sigma_1,\Sigma_2)$.
It may also be natural to consider three generations 
$(\Sigma_1,\Sigma_2,\Sigma_3)$, in line with the leptons and quarks of the Standard Model.
Since one ${\BS 2\BS 4}$ field carries the $SU(5)$ Dynkin index $N_5=5$, two $\Sigma_i$ contribute $N_5=10$ and three $\Sigma_i$ contribute $N_5=15$.
A larger Dynkin index gives a larger impact on the renormalization group flow.
The gauge coupling unification is maintained at the GUT scale as the messengers $\Sigma_i$ are in complete $SU(5)$ representations.
The magnitude of the unified coupling, in contrast, varies depending upon the number and the mass scale of the messengers.
As we will be interested in cosmological scenarios including inflation, theoretical consistency requires that the gauge couplings are perturbative beyond the unification scale, up to the scale where inflation takes place.

Solving the one loop renormalization group equation, the $SU(3)_c$ gauge coupling $g_3$ at the GUT scale $M_U$ is
\begin{align}
  \alpha_3^{-1}(M_U)=&
  \alpha_3^{-1}(M_t)+\frac{7}{2\pi}\ln\frac{M_S}{M_t}\crcr
  &+\frac{3}{2\pi}\ln\frac{M}{M_S}
  +\frac{3-N_5}{2\pi}\ln\frac{M_U}{M},
\end{align}
where $\alpha_3\equiv g_3^2/4\pi$, $M_t$ is the top quark mass, 
$M_S$ is the soft mass and $M$ is the messenger mass.
The flows of the $SU(2)_L$ and $U(1)_Y$ couplings are obtained similarly.
Above the GUT scale, the gauge couplings are assumed to be unified.
The beta function for the unified coupling $g_5$ turns out to be the same as that for the $g_3$ between the messenger mass scale $M$ and the GUT scale $M_U$.
Thus the unified gauge coupling at the inflation scale $M_{\rm inf}$ is
\begin{align}
  \alpha_5^{-1}(M_{\rm inf})=&
  \alpha_5^{-1}(M_U)+\frac{3-N_5}{2\pi}\ln\frac{M_{\rm inf}}{M_U},
\end{align}
where $\alpha_5=g_5^2/4\pi$. 
For larger $N_5$, the unified gauge coupling tends to diverge ($\alpha_5^{-1}$ hits zero) at a lower energy scale.
For a given scale of inflation, say $M_{\rm inf}\sim 10\, M_{\rm P}$ where $M_{\rm P}= 2.44\times 10^{18}$ GeV is the reduced Planck mass, we set the perturbativity condition for the gauge coupling up to inflation to be $\alpha_5(M_{\rm inf})\leq 1$.
This condition gives a lower bound on the messenger mass $M$.
With the typical soft mass $M_S=10$ TeV which is relevant for our scenario (see Table~\ref{tab:table1}), the bounds on the messenger mass are\footnote{We used the fitting formula \cite{Buttazzo:2013uya} to obtain the boundary conditions at $M_t$.}
\begin{align}\label{eqn:pertbounds}
  & M\geq 3.21\times 10^{11}\;\text{GeV with two }\Sigma_i,\crcr
  & M\geq 1.36\times 10^{14}\;\text{GeV with three }\Sigma_i.
\end{align}
These are strong constraints.
The three $\Sigma_i$ case of the $\nu$-GMSB model turns out to be incompatible with the lepton flavor violation bounds (see the next section).

\section{Lepton flavor violation}\label{sec:LFV}

In the type III seesaw $\nu$-GMSB model, the messenger fields $\Sigma_i$ and the matter fields $\overline F_i$ are coupled through the Dirac Yukawa coupling, see Eq.~\eqref{eqn:Wfull}, in addition to the gauge interactions.
This feature, called {\em Yukawa mediation}, is flavor dependent and the magnitude of the Yukawa coupling is constrained by experimental bounds on lepton flavor violating interactions.
The constraints on the Yukawa coupling, in turn, give an upper bound on the messenger mass $M$ via the seesaw formula \eqref{eqn:seesaw} in order to give the neutrino mass of the eV range which is natural for the neutrino oscillation data.

The lepton flavor violating decay rate is approximated by the formula \cite{Borzumati:1986qx,Hisano:1995cp}
\begin{align}
  \Gamma(\ell_i\to\ell_j\gamma)
  \sim\frac{\alpha_{\rm em}}{4}
  m_{\ell_i}^5\times\frac{\alpha_2^2}{16\pi^2}
  \frac{|(\Delta m_{\widetilde\ell}^2)_{ij}|^2}
  {m_{\widetilde\ell}^8}\tan^2\beta,
\end{align}
where
$\alpha_{\rm em}={e^2}/{4\pi}$, $\alpha_2={g_2^2}/{4\pi}$ for the electromagnetic and $SU(2)_L$ couplings, $m_{\widetilde\ell}$ is the flavor-diagonal soft mass arising from gauge mediation, and 
\begin{align}
  (\Delta m_{\widetilde\ell}^2)_{ij}\sim m_{\widetilde\ell}^2\frac{(Y_D^\dagger Y_D)_{ij}}{g_2^2},
\end{align}
are the left-handed slepton squared masses evaluated at the messenger scale.

In the model with two messenger fields $\Sigma_1, \Sigma_2$, 
the $Y_D^\dagger Y_D$ matrix elements are evaluated as\footnote{We chose the orthogonal matrix $R$ parametrizing the loose degrees of freedom to be unity; see Appendix~\ref{sec:A1}.}
\begin{align}\label{eqn:YDdYDNH}
& (Y_D^\dagger Y_D)^{\rm NH}_{ij}\crcr
&=\left(
\begin{array}{ccc}
 0.124 & 0.0162+0.155\, i & -0.164+0.142\, i \\
 0.0162-0.155\, i & 0.996 & 0.718-0.0178\, i \\
 -0.164-0.142\, i & 0.718+0.0178\, i & 0.849
\end{array}
\right)\crcr
&\qquad\times 10^{-3}\times\left(\frac{M}{10^{12}\;{\rm GeV}}\right)
\end{align}
for the normal mass hierarchy and  
\begin{align}\label{eqn:YDdYDIH}
& (Y_D^\dagger Y_D)^{\rm IH}_{ij}\crcr
&\!\!\!\!\!=\left(
\begin{array}{ccc}
 1.631 & 0.0844 -0.163\, i & 0.0613 -0.148 i \\
 0.0844 +0.163 i & 0.780 & -0.817 -0.00155 i \\
 0.0613 +0.148 i & -0.817 +0.00155 i & 0.933 \\
\end{array}
\right)\crcr
&\qquad\times 10^{-3}\times\left(\frac{M}{10^{12}\;{\rm GeV}}\right)
\end{align}
for the inverted mass hierarchy.
Our conventions of the neutrino mass matrix are summarized in Appendix~\ref{sec:A1}.
In Fig.~\ref{fig:LFV} we show the branching ratio ${\rm BR}(\mu\to e\,\gamma)$ for the normal and inverted mass hierarchies, together with the current experimental bound from the MEG experiments \cite{TheMEG:2016wtm}.
For the normal and inverted mass hierarchies, the upper bounds on the messenger mass $M$ are
\begin{align}
  & M\leq 2.28\times 10^{12}\;\text{GeV for NH},\crcr
  & M\leq 1.94\times 10^{12}\;\text{GeV for IH}.
\end{align}
Thus there is a mass range that satisfies the lepton flavor violation bounds and the perturbativity bounds \eqref{eqn:pertbounds}.

The model with three messengers is less predictive due to the extra ambiguities of the nonzero lightest neutrino mass.
For both normal and inverted mass hierarchies, the constraints on $M$ from lepton flavor violation are known to be more stringent than the two messenger case \cite{Mohapatra:2008wx}, and we see from \eqref{eqn:pertbounds} that there is no room for $M$ that satisfies both constraints.
We thus conclude that the three messenger model is not a viable option anymore.

\begin{figure}
\includegraphics[width=85mm]{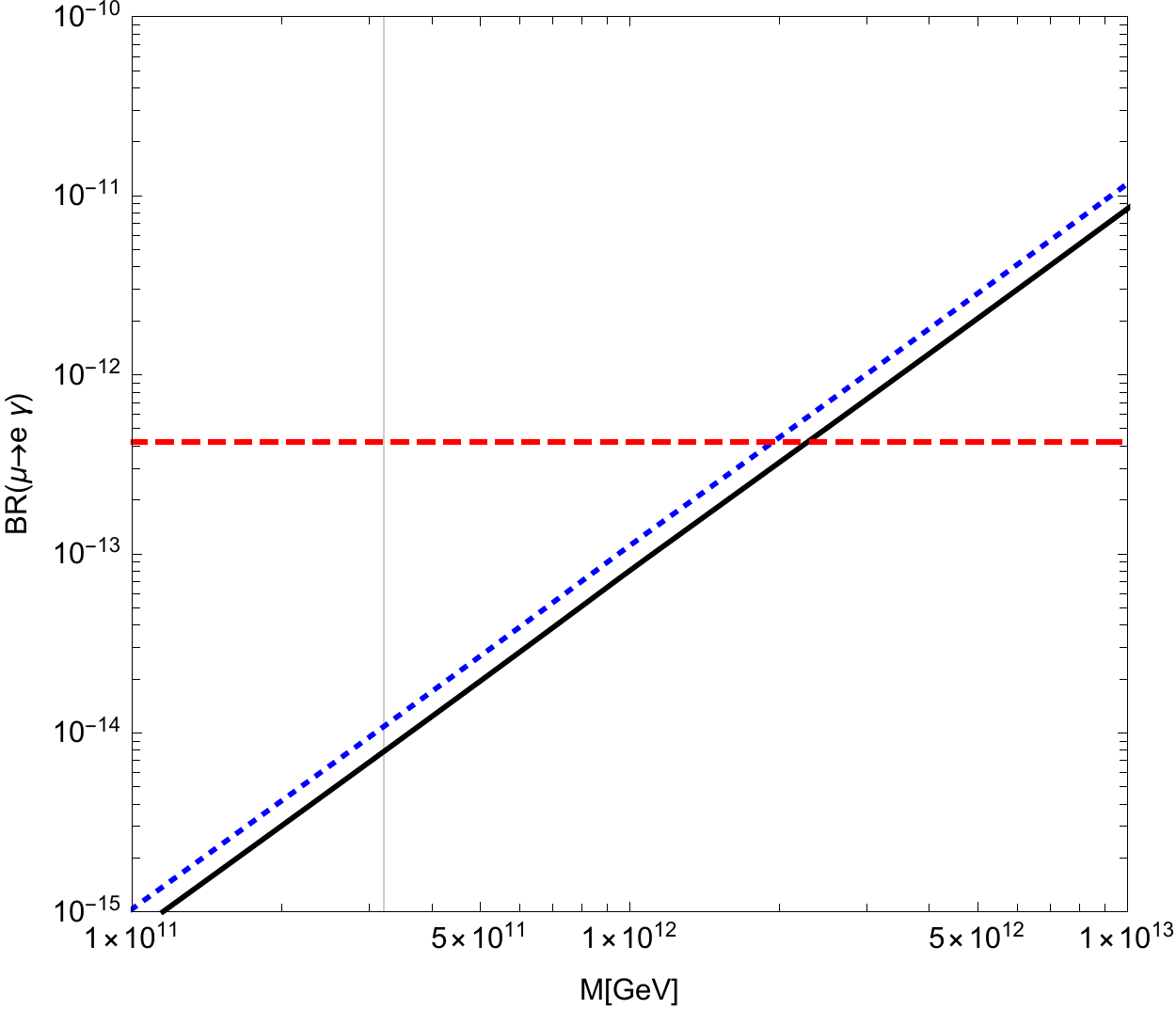}
\caption{
\label{fig:LFV}
Lepton flavor violating $\mu\to e\,\gamma$ branching ratio  for the normal mass hierarchy (the solid black line) and the inverted mass hierarchy (the dotted blue line).
The red dashed line indicates the experimental lower bound $4.2\times 10^{-13}$ for the branching ratio \cite{TheMEG:2016wtm}.
The thin vertical line is the lower bound on $M$ from the gauge coupling perturbativity (Sec.~\ref{sec:perturbativity}).
}
\end{figure}

\section{Inflation}\label{sec:inflation}

The mass spectrum of the GMSB scenario is determined by parameters
\begin{align}
  N_5,\; M,\; \Lambda\equiv\frac{yF_S}{M},\;\tan\beta,\;
  {\rm sign}\,\mu.
\end{align}
We fix $N_5=10$ as we will be interested in the two messenger model, and for the benchmark we choose $\tan\beta=10$ and ${\rm sign}\,\mu=+$.
The parameter $\Lambda$ is adjusted so that the predicted Higgs boson mass becomes 125.1 GeV \cite{Zyla:2020zbs}.
Table~\ref{tab:table1} shows the mass spectrum computed by the {\sc softsusy} 4.1.10 package \cite{Allanach:2001kg}, with the remaining parameter (the messenger mass) chosen to be $M=10^{11}, 10^{12}, 10^{13}$ GeV.
In this scenario, the lightest supersymmetric particle (LSP) is the gravitino and the next lightest supersymmetric particle (NLSP) is the stau.

After inflation, the gravitino is thermally produced and its abundance is given by \cite{Bolz:1998ek,Bolz:2000fu,Eberl:2020fml}
\begin{align}\label{eqn:Omega32}
  \Omega_{3/2} h^2 \sim 0.3\times \left(\frac{T_{\rm rh}}{10^8\;{\rm GeV}}\right)
  \left(\frac{1\;{\rm GeV}}{m_{3/2}}\right)\left(\frac{M_3}{1\;{\rm TeV}}\right)^2,
\end{align}
where $m_{3/2}=F_S/(\sqrt 3 M_{\rm P})$ is the gravitino mass and $M_3$ is the gluino ($\widetilde g$) mass, which is $M_3\sim 10$ TeV in the parameter range of our interest (Table \ref{tab:table1}).
The reheating temperature $T_{\rm rh}$ may be evaluated once the decay channels of the inflaton are specified.
Then the condition that the relic gravitino constitutes the major component of the dark matter together with the constraints by successful big bang nucleosynthesis (BBN) gives a consistency check for the scenario of inflation.

To construct a model of inflation, we use supergravity embedding of the $\nu$-GMSB model.
We use the superconformal framework \cite{Kaku:1978nz,Siegel:1978mj,Cremmer:1982en,Ferrara:1983dh,Kugo:1982mr,Kugo:1982cu,Kugo:1983mv}, by which the difficulty known as the eta problem is avoided and the observationally favored nonminimally coupled type of inflaton Lagrangian is naturally obtained.
See \cite{Ferrara:2010in,Arai:2011nq,Pallis:2011gr,Arai:2011aa,Arai:2012em,Arai:2013vaa,Kawai:2014doa,Kawai:2014gqa,Kawai:2015ryj,Leontaris:2016jty,Okada:2017rbf} for similar construction of inflationary models.
The resulting supergravity model has multidimensional moduli space, and one may consider different scenarios of cosmic inflation depending on along which flat direction inflation is assumed to take place. 
This ambiguity is attributed to the arbitrariness of the initial conditions and the non-uniqueness of the K\"ahler potential.
Below we present two case studies.

\subsection{Messenger inflation}\label{sec:MessInf}

One possible trajectory of inflaton is along the direction of the messenger fields \cite{Kawai:2021tzc} $\Sigma_i$. 
The ${\BS 2\BS 4}$ fields can be decomposed into components $\Sigma_j=\Sigma_j^aT^a$.
Let us suppose that $\Sigma_i^a$ has a large initial expectation value while the expectation values of the other fields are negligible.
Then the dynamics of inflation are dictated by the superpotential
\begin{align}
  W_{\rm inf} = \half(yS+M)(\Sigma_i^a)^2.
\end{align}
Assuming the K\"{a}hler potential 
\begin{align}
  {\C K}=-3M_{\rm P}^2+|\Sigma_i^a|^2+|S|^2-\frac 34\gamma \left\{(\Sigma_i^a)^2 + (\overline{\Sigma_i^a})^2\right\}+\cdots
\end{align}
in the superconformal framework\footnote{
The ellipsis must include a higher order term of $S$ (such as $|S|^4$) that controls the runaway behavior of the $S$ field; see  
\cite{Ferrara:2010in,Ferrara:2010yw}.} 
where $\gamma$ is a real parameter and writing the scalar part of the messenger as
\begin{align}
  \Sigma_i^a=\frac{1}{\sqrt 2}\varphi,
\end{align}
we may derive the scalar part of the supergravity action
\begin{align}\label{eqn:SJ}
  S_{\rm scalar}=\int d^4 x\sqrt{-g}\left[
\frac{M_{\rm P}^2+\xi\varphi^2}{2} R
-\half(\partial\varphi)^2
-V_{\rm J}(\varphi)\right],
\end{align}
where $R$ is the scalar curvature of the spacetime, $\xi\equiv\frac{\gamma}{4}-\frac 16$ and
\begin{align}\label{eqn:VJ}
V_{\rm J}(\varphi)=&\;
  \frac{y^2}{16}\varphi^4
  +M^2\varphi^2\frac{4M_{\rm P}^2-\left(\frac 16+2\gamma\right)\varphi^2}{8M_{\rm P}^2-\gamma(2-3\gamma)\varphi^2}.
\end{align}
The action \eqref{eqn:SJ} is in the Jordan frame.
It is transformed into the Einstein frame in which the scalar potential is
\begin{align}
  V_{\rm E}(\varphi)=\frac{y^2}{16}\frac{\varphi^4}{(M_{\rm P}^2+\xi\varphi^2)^2},
\end{align}
where the second term of \eqref{eqn:VJ} has been dropped since
$|y\varphi|\gg M$.
In the Einstein frame, the field $\varphi$ has a noncanonical kinetic term. 
The canonically normalized field $\widehat\varphi$ in the Einstein frame is written
\begin{align}
  \widehat\varphi
  =\int\frac{\sqrt{M_{\rm P}^2+\xi\varphi^2+6\xi^2\varphi^2}}{M_{\rm P}^2+\xi\varphi^2}d\varphi.
\end{align}
The model then reduces to the nonminimal $\varphi^4$ model that has been studied extensively \cite{Okada:2010jf}.
The prediction of the cosmic microwave background (CMB) spectrum is obtained in the standard slow roll paradigm.

Since the inflaton $\Sigma_i$ and the Standard Model Higgs field are directly coupled via the first term of \eqref{eqn:W}, the decay channel of the inflaton into the Standard Model particles is dominantly through the Higgs field.
Parametrizing the $SU(2)_L$ triplet $({\BS 1},{\BS 3},0)$ of \eqref{eqn:MMdecomp} as
\begin{align}
  \Sigma_{Ti}=\half\left(\begin{array}{cc}
    \Sigma_{Ti}^0 & \sqrt 2\Sigma_{Ti}^+ \\
    \sqrt 2\Sigma_{Ti}^- & -\Sigma_{Ti}^0
  \end{array}\right),
\end{align}
the part of the superpotential responsible for reheating is written
\begin{align}
  W_{\rm rh}= -Y_D^{ij}\nu_j\Sigma_{Ti}^0H_0.
\end{align}
The decay of the inflaton is through the interaction
\begin{align}
  {\C L}\supset-\frac{1}{\sqrt 2} Y_D^{ij}\nu_j\Sigma_{Ti}^{(\rm Re)}\widetilde H_0+ \text{H.c.}
\end{align}
where $\Sigma_{Ti}^{(\rm Re)}$ is the real part of $\Sigma_{Ti}$.
The decay width is
\begin{align}
  \Gamma(\Sigma_{Ti}^{(\rm Re)}\to\nu_j\widetilde H_0+\text{H.c.})
  =\frac{M}{16\pi}(Y_DY_D^\dagger)_{jj}
\end{align}
and assuming the perturbative reheating scenario the reheating temperature is evaluated as
\begin{align}\label{eqn:Treh}
  T_{\rm rh}\sim \sqrt{M_{\rm P}\Gamma}.
\end{align}
In the minimal seesaw model, the product of the Dirac Yukawa matrix is given by (see Appendix~\ref{sec:A1})
\begin{align}
  (Y_DY_D^\dagger)_{ij}=\left\{
  \begin{array}{ll}
    \frac{M}{v_u^2}{\rm diag}(m_2,m_3) & \text{Minimal seesaw, NH}\\\\
    \frac{M}{v_u^2}{\rm diag}(m_1,m_2) & \text{Minimal seesaw, IH}
  \end{array}\right.
\end{align}
and the smallest possible element is $(Y_DY_D^\dagger)_{11}$ in the case of the normal mass hierarchy, giving the minimal decay rate
\begin{align}
  \Gamma_{\rm min}=\frac{M}{16\pi}\frac{M}{v_u^2}m_2.
\end{align}
Using $m_2^2 = \Delta m_{21}^2=7.53\times 10^{-5}\;{\rm eV}^2$, the lower bound of the reheating temperature is then evaluated in the perturbative reheating scenario\footnote{
Possible nonlinear effects \cite{Dolgov:1989us,Traschen:1990sw,Kawai:2015lja} typically lead to higher reheating temperature.}
as
\begin{align}
  T_{\rm rh}\gtrsim 10^{13.5}\left(\frac{M}{10^{13}\,\text{GeV}}\right)\;\text{GeV}.
\end{align}
Thus, for the messenger mass of $M\sim 10^{13}\;\text{GeV}$, the thermally produced gravitino \eqref{eqn:Omega32} leads to overclosure of the Universe (the gravitino problem).

In the three messenger model, we have
\begin{align}
  (Y_DY_D^\dagger)_{ij}=\frac{M}{v_u^2}{\rm diag}(m_1,m_2,m_3)
\end{align}
and since $m_1$ is experimentally unconstrained, the lower bound of the reheating temperature may be evaluated for 
$(Y_D^\dagger Y_D)_{11}$ in the normal mass hierarchy case to be
\begin{align}
  T_{\rm rh}\sim 10^{6}
  \left(\frac{M}{10^{13}\,\text{GeV}}\right)
  \left(\frac{m_1}{10^{-17}\,\text{eV}}\right)
  \;\text{GeV}.
\end{align}
However, as discussed in Sec.~\ref{sec:LFV}, the perturbativity bound and the lepton flavor violation bound are not simultaneously satisfied in this case.
We thus have to conclude that the messenger inflation model is not phenomenologically viable.

\subsection{$L$-$H_u$ inflation}\label{sec:LHu}

Alternatively, one may consider a scenario in which inflation takes place in a flat direction along $\overline F_i F_H$ of the first term in \eqref{eqn:W}.
In this case, the superpotential responsible for the inflationary dynamics is written
\begin{align}
  W_{\rm inf}&= 
  \sqrt\frac{5}{2}\left(\frac{-3}{\sqrt{60}}\right)Y_D^{ij}(-\nu_j H_u^0)\widehat S_i\crcr
  +&\sqrt\frac{5}{2}\left(\frac{1}{2}\right)Y_D^{ij}(\nu_j H_u^0)\widehat T_i
  +\half M (\widehat S_i\widehat S_i+\widehat T_i\widehat T_i),
\end{align}
where $\widehat S_i$ and $\widehat T_i$ are the singlet and the triplet of the decomposition \eqref{eqn:MMdecomp}.
Likewise in the messenger inflation case above, let us assume that the K\"{a}hler potential takes a noncanonical form,
\begin{align}
  {\C K} =& -3M_{\rm P}^2+|H_u^0|^2+\sum_i\Big\{|\nu_i|^2+|\widehat S_i|^2+|\widehat T_i|^2\crcr
  &-\frac{3}{2}\gamma\left(\nu_iH_u^0+\text{H.c.}\right)\Big\}+\cdots.
\end{align}
Then the scalar part of the supergravity action is found to be in the same form \eqref{eqn:SJ}, with the scalar potential now given by 
\begin{align}
  V_{\rm J}
  =(Y_D^\dagger Y_D)^{ij}(\widetilde\nu_i H_u^0)(\widetilde\nu_j H_u^0)^\dagger.
\end{align}
For simplicity, let us consider $i=j=1,2,3$.
Among them, the flattest direction is likely to be the inflaton trajectory.
From \eqref{eqn:YDdYDNH} and \eqref{eqn:YDdYDIH}), it is natural to choose the $i=j=1$ direction
[$(Y_D^\dagger Y_D)^{11}=1.24\times 10^{-16}\,M$ GeV] for the normal mass hierarchy,
and the $i=j=2$ direction
[$(Y_D^\dagger Y_D)^{22}=7.80\times 10^{-16}\,M$ GeV] for the inverted mass hierarchy.

In this model, one of the inflaton components is the Standard Model Higgs field and thus the reheating temperature is evaluated as
\begin{align}
  T_{\rm rh}\simeq 
  \left(\frac{90}{\pi^2 g_*}\right)^{1/4}\sqrt{M_{\rm P}\Gamma_h}
  \simeq 6.8\times 10^7\;{\rm GeV},
\end{align}
where 
$\Gamma_h\simeq 4.07\times 10^{-3}\;{\rm GeV}$
is the Higgs decay width and $g_*\simeq 200$ for the supersymmetric model.
Thus, taking into account the effects of the redshift between the onset of reheating and the complete thermalization, the reasonable reheating temperature would be $T_{\rm rh}\sim 10^7$ GeV.
Then, with $M_3\sim 10$ TeV and $m_{3/2}\sim 10$ GeV, the gravitino abundance will be
\begin{align}
  \Omega_{3/2}h^2\simeq 0.1,
\end{align}
that is, appropriate value for the dark matter of the Universe.
The lifetime of the stau is evaluated as
\begin{align}
  \tau_{\widetilde\tau}\equiv &\Gamma(\widetilde\tau\to\tau\widetilde G)^{-1}
  \simeq\frac{48\pi\, m_{3/2}^2M_{\rm P}^2}{m_{\widetilde\tau}^5}
  \left(1-\frac{m_{3/2}^2}{m_{\widetilde\tau}^2}\right)^{-4}\crcr
  \simeq & 5.9\times 10^{-1}\left(\frac{m_{3/2}}{1\;\text{GeV}}\right)^2
  \left(\frac{1\;\text{TeV}}{m_{\widetilde\tau}}\right)^5\;\text{sec.}
\end{align}
For the gravitino mass $m_{3/2}\sim 10$ GeV and the stau mass $m_{\widetilde\tau}\sim 1$ TeV, the stau is found to be somewhat long lived.
This model is nevertheless compatible with the BBN bounds, as we can see from Fig.~12 in Ref. \cite{Kawasaki:2008qe}.

Thus, this model of inflation is phenomenologically viable. 
For our benchmark values
$\Lambda = yF_S/M\simeq 10^5$ GeV, 
$m_{3/2} = F_S/(\sqrt 3 M_{\rm P})\simeq 10$ GeV and
$M\sim 10^{12}$ GeV, 
the hidden sector Yukawa coupling is
\begin{align}
  y = \frac{\Lambda M}{F_S}=\frac{\Lambda M}{\sqrt 3 M_{\rm P}m_{3/2}}\sim 0.01.
\end{align}
Thus $y$ is also in the perturbative regime.
The CMB spectrum of the nonminimally coupled $\varphi^4$ inflation model is controlled by the e-folding number as well as by the inflaton self coupling, which is $(Y_D^\dagger Y_D)^{jj}$ in our case.
It can be checked that the prediction of the model, with the benchmark parameter values, fits well with the recent observational constraints \cite{Akrami:2018odb}.
The reheating temperature is somewhat lower than what is typically required for leptogenesis; a natural origin of the baryon asymmetry in this model would be due to the Affleck-Dine mechanism.

\section{Final remarks}\label{sec:final}

In this paper, we reinvestigated the type III seesaw $\nu$-GMSB model using the updated data, including the 125.1 GeV Higgs boson mass and the neutrino oscillation precision measurements.
We also proposed possible scenarios of cosmic inflation and discussed their phenomenological consistency.
We saw that the parameter region has been considerably narrowed compared to the time when the model was proposed; 
the three messenger model has already been ruled out, and the two messenger model is viable only for the narrow range of the messenger mass
$3\times 10^{11}\;\text{GeV}\;\lesssim M\lesssim 2\times 10^{12}\;\text{GeV}$.
We studied two possible scenarios of cosmic inflation;
one in which inflation takes place along the messenger direction, and the other in which inflation is along the $L$-$H_u$ flat direction.
The former model predicts high reheating temperature and suffers from the gravitino problem; it is thus phenomenologically unacceptable.
The latter model, on the other hand, satisfies all consistency requirements and is considered viable.

Let us conclude by commenting on how the model can be tested in future experiments.
The sensitivity of the lepton flavor violating $\mu\to e\gamma$ branching ratio 
${\rm Br}(\mu\to e\gamma)$ is expected to be improved by the MEG II experiment \cite{MEGII:2018kmf} up to $\sim 6\times 10^{-14}$.
In our model, this will give the upper bound on the messenger mass (see Fig.~\ref{fig:LFV})
\begin{align}
  & M\leq 9\times 10^{11}\;\text{GeV for NH},\crcr
  & M\leq 7\times 10^{11}\;\text{GeV for IH}.
\end{align}
Thus, a considerable part of the presently allowed parameter space would be covered. 
%
The current LHC bounds on the long-lived stau is 405 GeV maximum exclusion at $c\tau_{\widetilde\tau} = 2$ cm, that is, the lifetime of about $\tau_{\widetilde\tau} = 0.1$ ns (CMS \cite{CMS:2021kdm}). 
The parameter region of our interest $m_{\widetilde\tau}\sim$ 1 TeV, $\tau_{\widetilde\tau}\sim$ 10 sec. is presently unconstrained, and it will not be accessible even in the high luminosity upgrade of the LHC. 
The stau in this parameter region may be searched in a future collider, such as the proposed 100 TeV collider at CERN, with an experiment dedicated to the search of long-lived charged particles like MoEDAL, by a program similar to the one discussed in \cite{Felea:2020cvf}.
%

\subsection*{Acknowledgments}
We acknowledge useful communications with Kazunori Kohri.
This work was supported in part by the National Research Foundation of Korea Grant-in-Aid for Scientific Research Grant No.
NRF-2018R1D1A1B07051127
(S.K) and by the United States Department of Energy Grant No. DE-SC0012447 (N.O.).

\bigskip

\appendix
\section{Neutrino mass matrix}\label{sec:A1}

The neutrino Dirac mass matrix may be parametrized as \cite{Casas:2001sr,Ibarra:2003up}
\begin{align}
  (m_D)^{ij}=y_D^{ij}\langle H_u^0\rangle
  =\sqrt{M_R}R \sqrt{D_\nu} U_{\rm MNS}^\dagger,
\end{align}
where $U_{\rm MNS}$ is the Maki-Nakagawa-Sakata lepton flavor mixing matrix ($s_{ij}$ and $c_{ij}$ stand for $\sin\theta_{ij}$ and $\cos\theta_{ij}$)
\begin{align}
  U_{\rm MNS}=&\left(
  \begin{matrix}
 1 & 0 & 0 \\
 0 & c_{23} & s_{12}\\
 0 & -s_{23} & c_{23}
\end{matrix}
\right)
\left(
  \begin{matrix}
 c_{13} & 0 & s_{13}\,e^{-i\delta_{\rm CP}} \\
 0 & 1 & 0\\
 -s_{13}\,e^{i\delta_{\rm CP}} & 0 & c_{13}
\end{matrix}
\right)\crcr
&\times
\left(
  \begin{matrix}
 c_{12} & s_{12} & 0 \\
 -s_{12} & c_{12} & 0\\
 0 & 0 & 1
\end{matrix}
\right).
\end{align}
The matrix $\sqrt{M_R}$ is diagonal, and in our $\nu$GMSB model in which the messenger masses are assume to be degenerate, $\sqrt{M_R}$ is simply $\sqrt M$ multiplied by a unit matrix.
The complex orthogonal matrix $R$ represents our ignorance about the neutrino mass parameters. It was set to be a unit matrix in our analysis of lepton flavor violation as we are only interested in order estimation.

In the case of 3 generation messengers ($\Sigma_1$, $\Sigma_2$, $\Sigma_3$), the matrix $\sqrt{D_\nu}$ is  
\begin{align}
  \sqrt{D_\nu}={\rm diag}(\sqrt{m_1},\sqrt{m_2},\sqrt{m_3}),
\end{align}
where
\begin{align}
  m_2=\sqrt{\Delta m_{21}^2+m_1^2},\quad
  m_3=\sqrt{\Delta m_{32}^2+\Delta m_{21}^2+m_1^2}
\end{align}
for the normal mass hierarchy and
\begin{align}
  m_1=\sqrt{|\Delta m_{32}^2|-\Delta m_{21}^2+m_3^2},\quad
  m_2=\sqrt{|\Delta m_{32}^2|+m_3^2}
\end{align}
for the inverted mass hierarchy.

In the minimal seesaw model, 
\begin{align}
  \sqrt{D_\nu^{\rm NH}}=\left(\begin{array}{ccc}
  0 & \sqrt{m_2} & 0 \\ 
  0 & 0 & \sqrt{m_3} 
  \end{array}\right),
\end{align}
where $m_2=\sqrt{\Delta m_{21}^2}$ and $m_3=\sqrt{\Delta m_{32}^2+\Delta m_{21}^2}$
for the normal mass hierarchy and
\begin{align}
  \sqrt{D_\nu^{\rm IH}}=\left(\begin{array}{ccc} 
  \sqrt{m_1} & 0 & 0 \\ 
  0 & \sqrt{m_2} & 0 \\ 
  \end{array}\right),
\end{align}
where $m_1=\sqrt{|\Delta m_{32}^2|-\Delta m_{21}^2}$, $m_2=\sqrt{|\Delta m_{32}^2|}$ for the inverted mass hierarchy.

We used the following numerical values \cite{Zyla:2020zbs} in this paper:
\begin{align}
  \sin^2\theta_{12}=0.307,\quad
  \sin^2\theta_{13}=2.20\times 10^{-2},\crcr
  \Delta m_{21}^2=7.53\times 10^{-5}\;{\rm eV}^2,\quad
  \delta_{\rm CP}=1.36\,\pi,
\end{align}
for both mass hierarchies, and
\begin{align}
  \sin^2\theta_{23}=0.546,\quad
  \Delta m_{32}^2=2.453\times 10^{-3}\;{\rm eV}^2,
\end{align}
for the normal mass hierarchy and
\begin{align}
  \sin^2\theta_{23}=0.539,\quad
  \Delta m_{32}^2=-2.524\times 10^{-3}\;{\rm eV}^2,
\end{align}
for the inverted mass hierarchy.


\input{nuGMSB_Inf_3.bbl}

\end{document}

%% file: nuGMSB_Inf_3.bbl
%

%% file: nuGMSB_Inf_3.bbl
\begin{thebibliography}{51}%
\makeatletter
\providecommand \@ifxundefined [1]{%
 \@ifx{#1\undefined}
}%
\providecommand \@ifnum [1]{%
 \ifnum #1\expandafter \@firstoftwo
 \else \expandafter \@secondoftwo
 \fi
}%
\providecommand \@ifx [1]{%
 \ifx #1\expandafter \@firstoftwo
 \else \expandafter \@secondoftwo
 \fi
}%
\providecommand \natexlab [1]{#1}%
\providecommand \enquote  [1]{``#1''}%
\providecommand \bibnamefont  [1]{#1}%
\providecommand \bibfnamefont [1]{#1}%
\providecommand \citenamefont [1]{#1}%
\providecommand \href@noop [0]{\@secondoftwo}%
\providecommand \href [0]{\begingroup \@sanitize@url \@href}%
\providecommand \@href[1]{\@@startlink{#1}\@@href}%
\providecommand \@@href[1]{\endgroup#1\@@endlink}%
\providecommand \@sanitize@url [0]{\catcode `\\12\catcode `\$12\catcode
  `\&12\catcode `\#12\catcode `\^12\catcode `\_12\catcode `\%12\relax}%
\providecommand \@@startlink[1]{}%
\providecommand \@@endlink[0]{}%
\providecommand \url  [0]{\begingroup\@sanitize@url \@url }%
\providecommand \@url [1]{\endgroup\@href {#1}{\urlprefix }}%
\providecommand \urlprefix  [0]{URL }%
\providecommand \Eprint [0]{\href }%
\providecommand \doibase [0]{https://doi.org/}%
\providecommand \selectlanguage [0]{\@gobble}%
\providecommand \bibinfo  [0]{\@secondoftwo}%
\providecommand \bibfield  [0]{\@secondoftwo}%
\providecommand \translation [1]{[#1]}%
\providecommand \BibitemOpen [0]{}%
\providecommand \bibitemStop [0]{}%
\providecommand \bibitemNoStop [0]{.\EOS\space}%
\providecommand \EOS [0]{\spacefactor3000\relax}%
\providecommand \BibitemShut  [1]{\csname bibitem#1\endcsname}%
\let\auto@bib@innerbib\@empty
\bibitem [{\citenamefont {Minkowski}(1977)}]{Minkowski:1977sc}%
  \BibitemOpen
  \bibfield  {author} {\bibinfo {author} {\bibfnamefont {P.}~\bibnamefont
  {Minkowski}},\ }\bibfield  {title} {\bibinfo {title} {{$\mu \to e\gamma$ at a
  Rate of One Out of $10^{9}$ Muon Decays?}},\ }\href
  {https://doi.org/10.1016/0370-2693(77)90435-X} {\bibfield  {journal}
  {\bibinfo  {journal} {Phys. Lett. B}\ }\textbf {\bibinfo {volume} {67}},\
  \bibinfo {pages} {421} (\bibinfo {year} {1977})}\BibitemShut {NoStop}%
\bibitem [{\citenamefont {Yanagida}(1979)}]{Yanagida:1979as}%
  \BibitemOpen
  \bibfield  {author} {\bibinfo {author} {\bibfnamefont {T.}~\bibnamefont
  {Yanagida}},\ }\bibfield  {title} {\bibinfo {title} {{Horizontal gauge
  symmetry and masses of neutrinos}},\ }\href@noop {} {\bibfield  {journal}
  {\bibinfo  {journal} {Conf. Proc. C}\ }\textbf {\bibinfo {volume}
  {7902131}},\ \bibinfo {pages} {95} (\bibinfo {year} {1979})}\BibitemShut
  {NoStop}%
\bibitem [{\citenamefont {Gell-Mann}\ \emph {et~al.}(1979)\citenamefont
  {Gell-Mann}, \citenamefont {Ramond},\ and\ \citenamefont
  {Slansky}}]{GellMann:1980vs}%
  \BibitemOpen
  \bibfield  {author} {\bibinfo {author} {\bibfnamefont {M.}~\bibnamefont
  {Gell-Mann}}, \bibinfo {author} {\bibfnamefont {P.}~\bibnamefont {Ramond}},\
  and\ \bibinfo {author} {\bibfnamefont {R.}~\bibnamefont {Slansky}},\
  }\bibfield  {title} {\bibinfo {title} {{Complex Spinors and Unified
  Theories}},\ }\href@noop {} {\bibfield  {journal} {\bibinfo  {journal} {Conf.
  Proc. C}\ }\textbf {\bibinfo {volume} {790927}},\ \bibinfo {pages} {315}
  (\bibinfo {year} {1979})},\ \Eprint {https://arxiv.org/abs/1306.4669}
  {arXiv:1306.4669 [hep-th]} \BibitemShut {NoStop}%
\bibitem [{\citenamefont {Mohapatra}\ and\ \citenamefont
  {Senjanovic}(1980)}]{Mohapatra:1979ia}%
  \BibitemOpen
  \bibfield  {author} {\bibinfo {author} {\bibfnamefont {R.~N.}\ \bibnamefont
  {Mohapatra}}\ and\ \bibinfo {author} {\bibfnamefont {G.}~\bibnamefont
  {Senjanovic}},\ }\bibfield  {title} {\bibinfo {title} {{Neutrino Mass and
  Spontaneous Parity Violation}},\ }\href
  {https://doi.org/10.1103/PhysRevLett.44.912} {\bibfield  {journal} {\bibinfo
  {journal} {Phys.Rev.Lett.}\ }\textbf {\bibinfo {volume} {44}},\ \bibinfo
  {pages} {912} (\bibinfo {year} {1980})}\BibitemShut {NoStop}%
\bibitem [{\citenamefont {Joaquim}\ and\ \citenamefont
  {Rossi}(2006)}]{Joaquim:2006uz}%
  \BibitemOpen
  \bibfield  {author} {\bibinfo {author} {\bibfnamefont {F.~R.}\ \bibnamefont
  {Joaquim}}\ and\ \bibinfo {author} {\bibfnamefont {A.}~\bibnamefont
  {Rossi}},\ }\bibfield  {title} {\bibinfo {title} {{Gauge and Yukawa mediated
  supersymmetry breaking in the triplet seesaw scenario}},\ }\href
  {https://doi.org/10.1103/PhysRevLett.97.181801} {\bibfield  {journal}
  {\bibinfo  {journal} {Phys. Rev. Lett.}\ }\textbf {\bibinfo {volume} {97}},\
  \bibinfo {pages} {181801} (\bibinfo {year} {2006})},\ \Eprint
  {https://arxiv.org/abs/hep-ph/0604083} {arXiv:hep-ph/0604083} \BibitemShut
  {NoStop}%
\bibitem [{\citenamefont {Joaquim}\ and\ \citenamefont
  {Rossi}(2007)}]{Joaquim:2006mn}%
  \BibitemOpen
  \bibfield  {author} {\bibinfo {author} {\bibfnamefont {F.~R.}\ \bibnamefont
  {Joaquim}}\ and\ \bibinfo {author} {\bibfnamefont {A.}~\bibnamefont
  {Rossi}},\ }\bibfield  {title} {\bibinfo {title} {{Phenomenology of the
  triplet seesaw mechanism with Gauge and Yukawa mediation of SUSY breaking}},\
  }\href {https://doi.org/10.1016/j.nuclphysb.2006.11.030} {\bibfield
  {journal} {\bibinfo  {journal} {Nucl. Phys. B}\ }\textbf {\bibinfo {volume}
  {765}},\ \bibinfo {pages} {71} (\bibinfo {year} {2007})},\ \Eprint
  {https://arxiv.org/abs/hep-ph/0607298} {arXiv:hep-ph/0607298} \BibitemShut
  {NoStop}%
\bibitem [{\citenamefont {Mohapatra}\ \emph
  {et~al.}(2008{\natexlab{a}})\citenamefont {Mohapatra}, \citenamefont
  {Okada},\ and\ \citenamefont {Yu}}]{Mohapatra:2007js}%
  \BibitemOpen
  \bibfield  {author} {\bibinfo {author} {\bibfnamefont {R.~N.}\ \bibnamefont
  {Mohapatra}}, \bibinfo {author} {\bibfnamefont {N.}~\bibnamefont {Okada}},\
  and\ \bibinfo {author} {\bibfnamefont {H.-B.}\ \bibnamefont {Yu}},\
  }\bibfield  {title} {\bibinfo {title} {{Supersymmetry breaking by type II
  seesaw assisted anomaly mediation}},\ }\href
  {https://doi.org/10.1103/PhysRevD.77.115017} {\bibfield  {journal} {\bibinfo
  {journal} {Phys. Rev. D}\ }\textbf {\bibinfo {volume} {77}},\ \bibinfo
  {pages} {115017} (\bibinfo {year} {2008}{\natexlab{a}})},\ \Eprint
  {https://arxiv.org/abs/0711.0956} {arXiv:0711.0956 [hep-ph]} \BibitemShut
  {NoStop}%
\bibitem [{\citenamefont {Mohapatra}\ \emph
  {et~al.}(2008{\natexlab{b}})\citenamefont {Mohapatra}, \citenamefont
  {Setzer},\ and\ \citenamefont {Spinner}}]{Mohapatra:2008gz}%
  \BibitemOpen
  \bibfield  {author} {\bibinfo {author} {\bibfnamefont {R.~N.}\ \bibnamefont
  {Mohapatra}}, \bibinfo {author} {\bibfnamefont {N.}~\bibnamefont {Setzer}},\
  and\ \bibinfo {author} {\bibfnamefont {S.}~\bibnamefont {Spinner}},\
  }\bibfield  {title} {\bibinfo {title} {{Seesaw Extended MSSM and Anomaly
  Mediation without Tachyonic Sleptons}},\ }\href
  {https://doi.org/10.1088/1126-6708/2008/04/091} {\bibfield  {journal}
  {\bibinfo  {journal} {JHEP}\ }\textbf {\bibinfo {volume} {04}},\ \bibinfo
  {pages} {091}},\ \Eprint {https://arxiv.org/abs/0802.1208} {arXiv:0802.1208
  [hep-ph]} \BibitemShut {NoStop}%
\bibitem [{\citenamefont {Mohapatra}\ \emph
  {et~al.}(2008{\natexlab{c}})\citenamefont {Mohapatra}, \citenamefont
  {Okada},\ and\ \citenamefont {Yu}}]{Mohapatra:2008wx}%
  \BibitemOpen
  \bibfield  {author} {\bibinfo {author} {\bibfnamefont {R.~N.}\ \bibnamefont
  {Mohapatra}}, \bibinfo {author} {\bibfnamefont {N.}~\bibnamefont {Okada}},\
  and\ \bibinfo {author} {\bibfnamefont {H.-B.}\ \bibnamefont {Yu}},\
  }\bibfield  {title} {\bibinfo {title} {{nu-GMSB with Type III Seesaw and
  Phenomenology}},\ }\href {https://doi.org/10.1103/PhysRevD.78.075011}
  {\bibfield  {journal} {\bibinfo  {journal} {Phys. Rev. D}\ }\textbf {\bibinfo
  {volume} {78}},\ \bibinfo {pages} {075011} (\bibinfo {year}
  {2008}{\natexlab{c}})},\ \Eprint {https://arxiv.org/abs/0807.4524}
  {arXiv:0807.4524 [hep-ph]} \BibitemShut {NoStop}%
\bibitem [{\citenamefont {Arai}\ \emph {et~al.}(2010)\citenamefont {Arai},
  \citenamefont {Kawai},\ and\ \citenamefont {Okada}}]{Arai:2010ds}%
  \BibitemOpen
  \bibfield  {author} {\bibinfo {author} {\bibfnamefont {M.}~\bibnamefont
  {Arai}}, \bibinfo {author} {\bibfnamefont {S.}~\bibnamefont {Kawai}},\ and\
  \bibinfo {author} {\bibfnamefont {N.}~\bibnamefont {Okada}},\ }\bibfield
  {title} {\bibinfo {title} {{A Gauge mediation scenario with hidden sector
  renormalization in MSSM}},\ }\href
  {https://doi.org/10.1103/PhysRevD.81.035022} {\bibfield  {journal} {\bibinfo
  {journal} {Phys.Rev.}\ }\textbf {\bibinfo {volume} {D81}},\ \bibinfo {pages}
  {035022} (\bibinfo {year} {2010})},\ \Eprint
  {https://arxiv.org/abs/1001.1509} {arXiv:1001.1509 [hep-ph]} \BibitemShut
  {NoStop}%
\bibitem [{\citenamefont {Arai}\ \emph
  {et~al.}(2011{\natexlab{a}})\citenamefont {Arai}, \citenamefont {Kawai},\
  and\ \citenamefont {Okada}}]{Arai:2010qe}%
  \BibitemOpen
  \bibfield  {author} {\bibinfo {author} {\bibfnamefont {M.}~\bibnamefont
  {Arai}}, \bibinfo {author} {\bibfnamefont {S.}~\bibnamefont {Kawai}},\ and\
  \bibinfo {author} {\bibfnamefont {N.}~\bibnamefont {Okada}},\ }\bibfield
  {title} {\bibinfo {title} {{Renormalization effects on the MSSM from a
  calculable model of a strongly coupled hidden sector}},\ }\href
  {https://doi.org/10.1103/PhysRevD.84.075002} {\bibfield  {journal} {\bibinfo
  {journal} {Phys.Rev.}\ }\textbf {\bibinfo {volume} {D84}},\ \bibinfo {pages}
  {075002} (\bibinfo {year} {2011}{\natexlab{a}})},\ \Eprint
  {https://arxiv.org/abs/1011.3998} {arXiv:1011.3998 [hep-ph]} \BibitemShut
  {NoStop}%
\bibitem [{\citenamefont {Allanach}(2002)}]{Allanach:2001kg}%
  \BibitemOpen
  \bibfield  {author} {\bibinfo {author} {\bibfnamefont {B.~C.}\ \bibnamefont
  {Allanach}},\ }\bibfield  {title} {\bibinfo {title} {{SOFTSUSY: a program for
  calculating supersymmetric spectra}},\ }\href
  {https://doi.org/10.1016/S0010-4655(01)00460-X} {\bibfield  {journal}
  {\bibinfo  {journal} {Comput. Phys. Commun.}\ }\textbf {\bibinfo {volume}
  {143}},\ \bibinfo {pages} {305} (\bibinfo {year} {2002})},\ \Eprint
  {https://arxiv.org/abs/hep-ph/0104145} {arXiv:hep-ph/0104145} \BibitemShut
  {NoStop}%
\bibitem [{\citenamefont {Buttazzo}\ \emph {et~al.}(2013)\citenamefont
  {Buttazzo}, \citenamefont {Degrassi}, \citenamefont {Giardino}, \citenamefont
  {Giudice}, \citenamefont {Sala} \emph {et~al.}}]{Buttazzo:2013uya}%
  \BibitemOpen
  \bibfield  {author} {\bibinfo {author} {\bibfnamefont {D.}~\bibnamefont
  {Buttazzo}}, \bibinfo {author} {\bibfnamefont {G.}~\bibnamefont {Degrassi}},
  \bibinfo {author} {\bibfnamefont {P.~P.}\ \bibnamefont {Giardino}}, \bibinfo
  {author} {\bibfnamefont {G.~F.}\ \bibnamefont {Giudice}}, \bibinfo {author}
  {\bibfnamefont {F.}~\bibnamefont {Sala}}, \emph {et~al.},\ }\bibfield
  {title} {\bibinfo {title} {{Investigating the near-criticality of the Higgs
  boson}},\ }\href {https://doi.org/10.1007/JHEP12(2013)089} {\bibfield
  {journal} {\bibinfo  {journal} {JHEP}\ }\textbf {\bibinfo {volume} {1312}},\
  \bibinfo {pages} {089}},\ \Eprint {https://arxiv.org/abs/1307.3536}
  {arXiv:1307.3536 [hep-ph]} \BibitemShut {NoStop}%
\bibitem [{\citenamefont {Borzumati}\ and\ \citenamefont
  {Masiero}(1986)}]{Borzumati:1986qx}%
  \BibitemOpen
  \bibfield  {author} {\bibinfo {author} {\bibfnamefont {F.}~\bibnamefont
  {Borzumati}}\ and\ \bibinfo {author} {\bibfnamefont {A.}~\bibnamefont
  {Masiero}},\ }\bibfield  {title} {\bibinfo {title} {{Large Muon and electron
  Number Violations in Supergravity Theories}},\ }\href
  {https://doi.org/10.1103/PhysRevLett.57.961} {\bibfield  {journal} {\bibinfo
  {journal} {Phys. Rev. Lett.}\ }\textbf {\bibinfo {volume} {57}},\ \bibinfo
  {pages} {961} (\bibinfo {year} {1986})}\BibitemShut {NoStop}%
\bibitem [{\citenamefont {Hisano}\ \emph {et~al.}(1996)\citenamefont {Hisano},
  \citenamefont {Moroi}, \citenamefont {Tobe},\ and\ \citenamefont
  {Yamaguchi}}]{Hisano:1995cp}%
  \BibitemOpen
  \bibfield  {author} {\bibinfo {author} {\bibfnamefont {J.}~\bibnamefont
  {Hisano}}, \bibinfo {author} {\bibfnamefont {T.}~\bibnamefont {Moroi}},
  \bibinfo {author} {\bibfnamefont {K.}~\bibnamefont {Tobe}},\ and\ \bibinfo
  {author} {\bibfnamefont {M.}~\bibnamefont {Yamaguchi}},\ }\bibfield  {title}
  {\bibinfo {title} {{Lepton flavor violation via right-handed neutrino Yukawa
  couplings in supersymmetric standard model}},\ }\href
  {https://doi.org/10.1103/PhysRevD.53.2442} {\bibfield  {journal} {\bibinfo
  {journal} {Phys. Rev. D}\ }\textbf {\bibinfo {volume} {53}},\ \bibinfo
  {pages} {2442} (\bibinfo {year} {1996})},\ \Eprint
  {https://arxiv.org/abs/hep-ph/9510309} {arXiv:hep-ph/9510309} \BibitemShut
  {NoStop}%
\bibitem [{\citenamefont {Baldini}\ \emph {et~al.}(2016)\citenamefont {Baldini}
  \emph {et~al.}}]{TheMEG:2016wtm}%
  \BibitemOpen
  \bibfield  {author} {\bibinfo {author} {\bibfnamefont {A.~M.}\ \bibnamefont
  {Baldini}} \emph {et~al.} (\bibinfo {collaboration} {MEG}),\ }\bibfield
  {title} {\bibinfo {title} {{Search for the lepton flavour violating decay
  $\mu ^+ \rightarrow \mathrm {e}^+ \gamma $ with the full dataset of the MEG
  experiment}},\ }\href {https://doi.org/10.1140/epjc/s10052-016-4271-x}
  {\bibfield  {journal} {\bibinfo  {journal} {Eur. Phys. J. C}\ }\textbf
  {\bibinfo {volume} {76}},\ \bibinfo {pages} {434} (\bibinfo {year} {2016})},\
  \Eprint {https://arxiv.org/abs/1605.05081} {arXiv:1605.05081 [hep-ex]}
  \BibitemShut {NoStop}%
\bibitem [{\citenamefont {Zyla}\ \emph {et~al.}(2020)\citenamefont {Zyla} \emph
  {et~al.}}]{Zyla:2020zbs}%
  \BibitemOpen
  \bibfield  {author} {\bibinfo {author} {\bibfnamefont {P.~A.}\ \bibnamefont
  {Zyla}} \emph {et~al.} (\bibinfo {collaboration} {Particle Data Group}),\
  }\bibfield  {title} {\bibinfo {title} {{Review of Particle Physics}},\ }\href
  {https://doi.org/10.1093/ptep/ptaa104} {\bibfield  {journal} {\bibinfo
  {journal} {PTEP}\ }\textbf {\bibinfo {volume} {2020}},\ \bibinfo {pages}
  {083C01} (\bibinfo {year} {2020})}\BibitemShut {NoStop}%
\bibitem [{\citenamefont {Bolz}\ \emph {et~al.}(1998)\citenamefont {Bolz},
  \citenamefont {Buchmuller},\ and\ \citenamefont {Plumacher}}]{Bolz:1998ek}%
  \BibitemOpen
  \bibfield  {author} {\bibinfo {author} {\bibfnamefont {M.}~\bibnamefont
  {Bolz}}, \bibinfo {author} {\bibfnamefont {W.}~\bibnamefont {Buchmuller}},\
  and\ \bibinfo {author} {\bibfnamefont {M.}~\bibnamefont {Plumacher}},\
  }\bibfield  {title} {\bibinfo {title} {{Baryon asymmetry and dark matter}},\
  }\href {https://doi.org/10.1016/S0370-2693(98)01342-2} {\bibfield  {journal}
  {\bibinfo  {journal} {Phys. Lett. B}\ }\textbf {\bibinfo {volume} {443}},\
  \bibinfo {pages} {209} (\bibinfo {year} {1998})},\ \Eprint
  {https://arxiv.org/abs/hep-ph/9809381} {arXiv:hep-ph/9809381} \BibitemShut
  {NoStop}%
\bibitem [{\citenamefont {Bolz}\ \emph {et~al.}(2001)\citenamefont {Bolz},
  \citenamefont {Brandenburg},\ and\ \citenamefont {Buchmuller}}]{Bolz:2000fu}%
  \BibitemOpen
  \bibfield  {author} {\bibinfo {author} {\bibfnamefont {M.}~\bibnamefont
  {Bolz}}, \bibinfo {author} {\bibfnamefont {A.}~\bibnamefont {Brandenburg}},\
  and\ \bibinfo {author} {\bibfnamefont {W.}~\bibnamefont {Buchmuller}},\
  }\bibfield  {title} {\bibinfo {title} {{Thermal production of gravitinos}},\
  }\href {https://doi.org/10.1016/S0550-3213(01)00132-8} {\bibfield  {journal}
  {\bibinfo  {journal} {Nucl.Phys.}\ }\textbf {\bibinfo {volume} {B606}},\
  \bibinfo {pages} {518} (\bibinfo {year} {2001})},\ \Eprint
  {https://arxiv.org/abs/hep-ph/0012052} {arXiv:hep-ph/0012052 [hep-ph]}
  \BibitemShut {NoStop}%
\bibitem [{\citenamefont {Eberl}\ \emph {et~al.}(2021)\citenamefont {Eberl},
  \citenamefont {Gialamas},\ and\ \citenamefont {Spanos}}]{Eberl:2020fml}%
  \BibitemOpen
  \bibfield  {author} {\bibinfo {author} {\bibfnamefont {H.}~\bibnamefont
  {Eberl}}, \bibinfo {author} {\bibfnamefont {I.~D.}\ \bibnamefont
  {Gialamas}},\ and\ \bibinfo {author} {\bibfnamefont {V.~C.}\ \bibnamefont
  {Spanos}},\ }\bibfield  {title} {\bibinfo {title} {{Gravitino thermal
  production revisited}},\ }\href {https://doi.org/10.1103/PhysRevD.103.075025}
  {\bibfield  {journal} {\bibinfo  {journal} {Phys. Rev. D}\ }\textbf {\bibinfo
  {volume} {103}},\ \bibinfo {pages} {075025} (\bibinfo {year} {2021})},\
  \Eprint {https://arxiv.org/abs/2010.14621} {arXiv:2010.14621 [hep-ph]}
  \BibitemShut {NoStop}%
\bibitem [{\citenamefont {Kaku}\ \emph {et~al.}(1978)\citenamefont {Kaku},
  \citenamefont {Townsend},\ and\ \citenamefont {van
  Nieuwenhuizen}}]{Kaku:1978nz}%
  \BibitemOpen
  \bibfield  {author} {\bibinfo {author} {\bibfnamefont {M.}~\bibnamefont
  {Kaku}}, \bibinfo {author} {\bibfnamefont {P.}~\bibnamefont {Townsend}},\
  and\ \bibinfo {author} {\bibfnamefont {P.}~\bibnamefont {van
  Nieuwenhuizen}},\ }\bibfield  {title} {\bibinfo {title} {{Properties of
  Conformal Supergravity}},\ }\href {https://doi.org/10.1103/PhysRevD.17.3179}
  {\bibfield  {journal} {\bibinfo  {journal} {Phys.Rev.}\ }\textbf {\bibinfo
  {volume} {D17}},\ \bibinfo {pages} {3179} (\bibinfo {year}
  {1978})}\BibitemShut {NoStop}%
\bibitem [{\citenamefont {Siegel}\ and\ \citenamefont
  {Gates}(1979)}]{Siegel:1978mj}%
  \BibitemOpen
  \bibfield  {author} {\bibinfo {author} {\bibfnamefont {W.}~\bibnamefont
  {Siegel}}\ and\ \bibinfo {author} {\bibfnamefont {J.}~\bibnamefont {Gates},
  \bibfnamefont {S.~James}},\ }\bibfield  {title} {\bibinfo {title}
  {{Superfield Supergravity}},\ }\href
  {https://doi.org/10.1016/0550-3213(79)90416-4} {\bibfield  {journal}
  {\bibinfo  {journal} {Nucl.Phys.}\ }\textbf {\bibinfo {volume} {B147}},\
  \bibinfo {pages} {77} (\bibinfo {year} {1979})}\BibitemShut {NoStop}%
\bibitem [{\citenamefont {Cremmer}\ \emph {et~al.}(1983)\citenamefont
  {Cremmer}, \citenamefont {Ferrara}, \citenamefont {Girardello},\ and\
  \citenamefont {Van~Proeyen}}]{Cremmer:1982en}%
  \BibitemOpen
  \bibfield  {author} {\bibinfo {author} {\bibfnamefont {E.}~\bibnamefont
  {Cremmer}}, \bibinfo {author} {\bibfnamefont {S.}~\bibnamefont {Ferrara}},
  \bibinfo {author} {\bibfnamefont {L.}~\bibnamefont {Girardello}},\ and\
  \bibinfo {author} {\bibfnamefont {A.}~\bibnamefont {Van~Proeyen}},\
  }\bibfield  {title} {\bibinfo {title} {{Yang-Mills Theories with Local
  Supersymmetry: Lagrangian, Transformation Laws and SuperHiggs Effect}},\
  }\href {https://doi.org/10.1016/0550-3213(83)90679-X} {\bibfield  {journal}
  {\bibinfo  {journal} {Nucl. Phys. B}\ }\textbf {\bibinfo {volume} {212}},\
  \bibinfo {pages} {413} (\bibinfo {year} {1983})}\BibitemShut {NoStop}%
\bibitem [{\citenamefont {Ferrara}\ \emph {et~al.}(1983)\citenamefont
  {Ferrara}, \citenamefont {Girardello}, \citenamefont {Kugo},\ and\
  \citenamefont {Van~Proeyen}}]{Ferrara:1983dh}%
  \BibitemOpen
  \bibfield  {author} {\bibinfo {author} {\bibfnamefont {S.}~\bibnamefont
  {Ferrara}}, \bibinfo {author} {\bibfnamefont {L.}~\bibnamefont {Girardello}},
  \bibinfo {author} {\bibfnamefont {T.}~\bibnamefont {Kugo}},\ and\ \bibinfo
  {author} {\bibfnamefont {A.}~\bibnamefont {Van~Proeyen}},\ }\bibfield
  {title} {\bibinfo {title} {{Relation Between Different Auxiliary Field
  Formulations of $N=1$ Supergravity Coupled to Matter}},\ }\href
  {https://doi.org/10.1016/0550-3213(83)90101-3} {\bibfield  {journal}
  {\bibinfo  {journal} {Nucl. Phys. B}\ }\textbf {\bibinfo {volume} {223}},\
  \bibinfo {pages} {191} (\bibinfo {year} {1983})}\BibitemShut {NoStop}%
\bibitem [{\citenamefont {Kugo}\ and\ \citenamefont
  {Uehara}(1983{\natexlab{a}})}]{Kugo:1982mr}%
  \BibitemOpen
  \bibfield  {author} {\bibinfo {author} {\bibfnamefont {T.}~\bibnamefont
  {Kugo}}\ and\ \bibinfo {author} {\bibfnamefont {S.}~\bibnamefont {Uehara}},\
  }\bibfield  {title} {\bibinfo {title} {{Improved Superconformal Gauge
  Conditions in the $N=1$ Supergravity \{Yang-Mills\} Matter System}},\ }\href
  {https://doi.org/10.1016/0550-3213(83)90612-0} {\bibfield  {journal}
  {\bibinfo  {journal} {Nucl. Phys. B}\ }\textbf {\bibinfo {volume} {222}},\
  \bibinfo {pages} {125} (\bibinfo {year} {1983}{\natexlab{a}})}\BibitemShut
  {NoStop}%
\bibitem [{\citenamefont {Kugo}\ and\ \citenamefont
  {Uehara}(1983{\natexlab{b}})}]{Kugo:1982cu}%
  \BibitemOpen
  \bibfield  {author} {\bibinfo {author} {\bibfnamefont {T.}~\bibnamefont
  {Kugo}}\ and\ \bibinfo {author} {\bibfnamefont {S.}~\bibnamefont {Uehara}},\
  }\bibfield  {title} {\bibinfo {title} {{Conformal and Poincare Tensor Calculi
  in $N=1$ Supergravity}},\ }\href
  {https://doi.org/10.1016/0550-3213(83)90463-7} {\bibfield  {journal}
  {\bibinfo  {journal} {Nucl. Phys. B}\ }\textbf {\bibinfo {volume} {226}},\
  \bibinfo {pages} {49} (\bibinfo {year} {1983}{\natexlab{b}})}\BibitemShut
  {NoStop}%
\bibitem [{\citenamefont {Kugo}\ and\ \citenamefont
  {Uehara}(1985)}]{Kugo:1983mv}%
  \BibitemOpen
  \bibfield  {author} {\bibinfo {author} {\bibfnamefont {T.}~\bibnamefont
  {Kugo}}\ and\ \bibinfo {author} {\bibfnamefont {S.}~\bibnamefont {Uehara}},\
  }\bibfield  {title} {\bibinfo {title} {{$N=1$ Superconformal Tensor Calculus:
  Multiplets With External Lorentz Indices and Spinor Derivative Operators}},\
  }\href {https://doi.org/10.1143/PTP.73.235} {\bibfield  {journal} {\bibinfo
  {journal} {Prog. Theor. Phys.}\ }\textbf {\bibinfo {volume} {73}},\ \bibinfo
  {pages} {235} (\bibinfo {year} {1985})}\BibitemShut {NoStop}%
\bibitem [{\citenamefont {Ferrara}\ \emph {et~al.}(2011)\citenamefont
  {Ferrara}, \citenamefont {Kallosh}, \citenamefont {Linde}, \citenamefont
  {Marrani},\ and\ \citenamefont {Van~Proeyen}}]{Ferrara:2010in}%
  \BibitemOpen
  \bibfield  {author} {\bibinfo {author} {\bibfnamefont {S.}~\bibnamefont
  {Ferrara}}, \bibinfo {author} {\bibfnamefont {R.}~\bibnamefont {Kallosh}},
  \bibinfo {author} {\bibfnamefont {A.}~\bibnamefont {Linde}}, \bibinfo
  {author} {\bibfnamefont {A.}~\bibnamefont {Marrani}},\ and\ \bibinfo {author}
  {\bibfnamefont {A.}~\bibnamefont {Van~Proeyen}},\ }\bibfield  {title}
  {\bibinfo {title} {{Superconformal Symmetry, NMSSM, and Inflation}},\ }\href
  {https://doi.org/10.1103/PhysRevD.83.025008} {\bibfield  {journal} {\bibinfo
  {journal} {Phys.Rev.}\ }\textbf {\bibinfo {volume} {D83}},\ \bibinfo {pages}
  {025008} (\bibinfo {year} {2011})},\ \Eprint
  {https://arxiv.org/abs/1008.2942} {arXiv:1008.2942 [hep-th]} \BibitemShut
  {NoStop}%
\bibitem [{\citenamefont {Arai}\ \emph
  {et~al.}(2011{\natexlab{b}})\citenamefont {Arai}, \citenamefont {Kawai},\
  and\ \citenamefont {Okada}}]{Arai:2011nq}%
  \BibitemOpen
  \bibfield  {author} {\bibinfo {author} {\bibfnamefont {M.}~\bibnamefont
  {Arai}}, \bibinfo {author} {\bibfnamefont {S.}~\bibnamefont {Kawai}},\ and\
  \bibinfo {author} {\bibfnamefont {N.}~\bibnamefont {Okada}},\ }\bibfield
  {title} {\bibinfo {title} {{Higgs inflation in minimal supersymmetric SU(5)
  GUT}},\ }\href {https://doi.org/10.1103/PhysRevD.84.123515} {\bibfield
  {journal} {\bibinfo  {journal} {Phys.Rev.}\ }\textbf {\bibinfo {volume}
  {D84}},\ \bibinfo {pages} {123515} (\bibinfo {year} {2011}{\natexlab{b}})},\
  \Eprint {https://arxiv.org/abs/1107.4767} {arXiv:1107.4767 [hep-ph]}
  \BibitemShut {NoStop}%
\bibitem [{\citenamefont {Pallis}\ and\ \citenamefont
  {Toumbas}(2011)}]{Pallis:2011gr}%
  \BibitemOpen
  \bibfield  {author} {\bibinfo {author} {\bibfnamefont {C.}~\bibnamefont
  {Pallis}}\ and\ \bibinfo {author} {\bibfnamefont {N.}~\bibnamefont
  {Toumbas}},\ }\bibfield  {title} {\bibinfo {title} {{Non-Minimal Higgs
  Inflation and non-Thermal Leptogenesis in A Supersymmetric Pati-Salam
  Model}},\ }\href {https://doi.org/10.1088/1475-7516/2011/12/002} {\bibfield
  {journal} {\bibinfo  {journal} {JCAP}\ }\textbf {\bibinfo {volume} {1112}},\
  \bibinfo {pages} {002}},\ \Eprint {https://arxiv.org/abs/1108.1771}
  {arXiv:1108.1771 [hep-ph]} \BibitemShut {NoStop}%
\bibitem [{\citenamefont {Arai}\ \emph {et~al.}(2012)\citenamefont {Arai},
  \citenamefont {Kawai},\ and\ \citenamefont {Okada}}]{Arai:2011aa}%
  \BibitemOpen
  \bibfield  {author} {\bibinfo {author} {\bibfnamefont {M.}~\bibnamefont
  {Arai}}, \bibinfo {author} {\bibfnamefont {S.}~\bibnamefont {Kawai}},\ and\
  \bibinfo {author} {\bibfnamefont {N.}~\bibnamefont {Okada}},\ }\bibfield
  {title} {\bibinfo {title} {{Supersymmetric standard model inflation in the
  Planck era}},\ }\href {https://doi.org/10.1103/PhysRevD.86.063507} {\bibfield
   {journal} {\bibinfo  {journal} {Phys.Rev.}\ }\textbf {\bibinfo {volume}
  {D86}},\ \bibinfo {pages} {063507} (\bibinfo {year} {2012})},\ \Eprint
  {https://arxiv.org/abs/1112.2391} {arXiv:1112.2391 [hep-ph]} \BibitemShut
  {NoStop}%
\bibitem [{\citenamefont {Arai}\ \emph {et~al.}(2013)\citenamefont {Arai},
  \citenamefont {Kawai},\ and\ \citenamefont {Okada}}]{Arai:2012em}%
  \BibitemOpen
  \bibfield  {author} {\bibinfo {author} {\bibfnamefont {M.}~\bibnamefont
  {Arai}}, \bibinfo {author} {\bibfnamefont {S.}~\bibnamefont {Kawai}},\ and\
  \bibinfo {author} {\bibfnamefont {N.}~\bibnamefont {Okada}},\ }\bibfield
  {title} {\bibinfo {title} {{Higgs-lepton inflation in the supersymmetric
  minimal seesaw model}},\ }\href {https://doi.org/10.1103/PhysRevD.87.065009}
  {\bibfield  {journal} {\bibinfo  {journal} {Phys.Rev.}\ }\textbf {\bibinfo
  {volume} {D87}},\ \bibinfo {pages} {065009} (\bibinfo {year} {2013})},\
  \Eprint {https://arxiv.org/abs/1212.6828} {arXiv:1212.6828 [hep-ph]}
  \BibitemShut {NoStop}%
\bibitem [{\citenamefont {Arai}\ \emph {et~al.}(2014)\citenamefont {Arai},
  \citenamefont {Kawai},\ and\ \citenamefont {Okada}}]{Arai:2013vaa}%
  \BibitemOpen
  \bibfield  {author} {\bibinfo {author} {\bibfnamefont {M.}~\bibnamefont
  {Arai}}, \bibinfo {author} {\bibfnamefont {S.}~\bibnamefont {Kawai}},\ and\
  \bibinfo {author} {\bibfnamefont {N.}~\bibnamefont {Okada}},\ }\bibfield
  {title} {\bibinfo {title} {{Supersymmetric B\ensuremath{-}L inflation near
  the conformal coupling}},\ }\href
  {https://doi.org/10.1016/j.physletb.2014.05.027} {\bibfield  {journal}
  {\bibinfo  {journal} {Phys. Lett. B}\ }\textbf {\bibinfo {volume} {734}},\
  \bibinfo {pages} {100} (\bibinfo {year} {2014})},\ \Eprint
  {https://arxiv.org/abs/1311.1317} {arXiv:1311.1317 [hep-ph]} \BibitemShut
  {NoStop}%
\bibitem [{\citenamefont {Kawai}\ and\ \citenamefont
  {Okada}(2014)}]{Kawai:2014doa}%
  \BibitemOpen
  \bibfield  {author} {\bibinfo {author} {\bibfnamefont {S.}~\bibnamefont
  {Kawai}}\ and\ \bibinfo {author} {\bibfnamefont {N.}~\bibnamefont {Okada}},\
  }\bibfield  {title} {\bibinfo {title} {{TeV scale seesaw from supersymmetric
  Higgs-lepton inflation and BICEP2}},\ }\href
  {https://doi.org/10.1016/j.physletb.2014.06.042} {\bibfield  {journal}
  {\bibinfo  {journal} {Phys. Lett.}\ }\textbf {\bibinfo {volume} {B735}},\
  \bibinfo {pages} {186} (\bibinfo {year} {2014})},\ \Eprint
  {https://arxiv.org/abs/1404.1450} {arXiv:1404.1450 [hep-ph]} \BibitemShut
  {NoStop}%
\bibitem [{\citenamefont {Kawai}\ and\ \citenamefont
  {Kim}(2015)}]{Kawai:2014gqa}%
  \BibitemOpen
  \bibfield  {author} {\bibinfo {author} {\bibfnamefont {S.}~\bibnamefont
  {Kawai}}\ and\ \bibinfo {author} {\bibfnamefont {J.}~\bibnamefont {Kim}},\
  }\bibfield  {title} {\bibinfo {title} {{Testing supersymmetric Higgs
  inflation with non-Gaussianity}},\ }\href
  {https://doi.org/10.1103/PhysRevD.91.045021} {\bibfield  {journal} {\bibinfo
  {journal} {Phys.Rev.}\ }\textbf {\bibinfo {volume} {D91}},\ \bibinfo {pages}
  {045021} (\bibinfo {year} {2015})},\ \Eprint
  {https://arxiv.org/abs/1411.5188} {arXiv:1411.5188 [hep-ph]} \BibitemShut
  {NoStop}%
\bibitem [{\citenamefont {Kawai}\ and\ \citenamefont
  {Kim}(2016)}]{Kawai:2015ryj}%
  \BibitemOpen
  \bibfield  {author} {\bibinfo {author} {\bibfnamefont {S.}~\bibnamefont
  {Kawai}}\ and\ \bibinfo {author} {\bibfnamefont {J.}~\bibnamefont {Kim}},\
  }\bibfield  {title} {\bibinfo {title} {{Multifield dynamics of supersymmetric
  Higgs inflation in SU(5) GUT}},\ }\href
  {https://doi.org/10.1103/PhysRevD.93.065023} {\bibfield  {journal} {\bibinfo
  {journal} {Phys. Rev. D}\ }\textbf {\bibinfo {volume} {93}},\ \bibinfo
  {pages} {065023} (\bibinfo {year} {2016})},\ \Eprint
  {https://arxiv.org/abs/1512.05861} {arXiv:1512.05861 [hep-ph]} \BibitemShut
  {NoStop}%
\bibitem [{\citenamefont {Leontaris}\ \emph {et~al.}(2017)\citenamefont
  {Leontaris}, \citenamefont {Okada},\ and\ \citenamefont
  {Shafi}}]{Leontaris:2016jty}%
  \BibitemOpen
  \bibfield  {author} {\bibinfo {author} {\bibfnamefont {G.~K.}\ \bibnamefont
  {Leontaris}}, \bibinfo {author} {\bibfnamefont {N.}~\bibnamefont {Okada}},\
  and\ \bibinfo {author} {\bibfnamefont {Q.}~\bibnamefont {Shafi}},\ }\bibfield
   {title} {\bibinfo {title} {{Non-minimal quartic inflation in supersymmetric
  $SO(10)$}},\ }\href {https://doi.org/10.1016/j.physletb.2016.12.038}
  {\bibfield  {journal} {\bibinfo  {journal} {Phys. Lett. B}\ }\textbf
  {\bibinfo {volume} {765}},\ \bibinfo {pages} {256} (\bibinfo {year}
  {2017})},\ \Eprint {https://arxiv.org/abs/1611.10196} {arXiv:1611.10196
  [hep-ph]} \BibitemShut {NoStop}%
\bibitem [{\citenamefont {Okada}\ and\ \citenamefont
  {Shafi}(2018)}]{Okada:2017rbf}%
  \BibitemOpen
  \bibfield  {author} {\bibinfo {author} {\bibfnamefont {N.}~\bibnamefont
  {Okada}}\ and\ \bibinfo {author} {\bibfnamefont {Q.}~\bibnamefont {Shafi}},\
  }\bibfield  {title} {\bibinfo {title} {{Gravity waves and gravitino dark
  matter in $\mu$-hybrid inflation}},\ }\href
  {https://doi.org/10.1016/j.physletb.2018.10.057} {\bibfield  {journal}
  {\bibinfo  {journal} {Phys. Lett. B}\ }\textbf {\bibinfo {volume} {787}},\
  \bibinfo {pages} {141} (\bibinfo {year} {2018})},\ \Eprint
  {https://arxiv.org/abs/1709.04610} {arXiv:1709.04610 [hep-ph]} \BibitemShut
  {NoStop}%
\bibitem [{\citenamefont {Kawai}\ and\ \citenamefont
  {Okada}(2021)}]{Kawai:2021tzc}%
  \BibitemOpen
  \bibfield  {author} {\bibinfo {author} {\bibfnamefont {S.}~\bibnamefont
  {Kawai}}\ and\ \bibinfo {author} {\bibfnamefont {N.}~\bibnamefont {Okada}},\
  }\bibfield  {title} {\bibinfo {title} {{Messenger inflation in gauge
  mediation and super-WIMP dark matter}},\ }\href
  {https://doi.org/10.1103/PhysRevD.104.083539} {\bibfield  {journal} {\bibinfo
   {journal} {Phys. Rev. D}\ }\textbf {\bibinfo {volume} {104}},\ \bibinfo
  {pages} {083539} (\bibinfo {year} {2021})},\ \Eprint
  {https://arxiv.org/abs/2103.11256} {arXiv:2103.11256 [hep-ph]} \BibitemShut
  {NoStop}%
\bibitem [{\citenamefont {Ferrara}\ \emph {et~al.}(2010)\citenamefont
  {Ferrara}, \citenamefont {Kallosh}, \citenamefont {Linde}, \citenamefont
  {Marrani},\ and\ \citenamefont {Van~Proeyen}}]{Ferrara:2010yw}%
  \BibitemOpen
  \bibfield  {author} {\bibinfo {author} {\bibfnamefont {S.}~\bibnamefont
  {Ferrara}}, \bibinfo {author} {\bibfnamefont {R.}~\bibnamefont {Kallosh}},
  \bibinfo {author} {\bibfnamefont {A.}~\bibnamefont {Linde}}, \bibinfo
  {author} {\bibfnamefont {A.}~\bibnamefont {Marrani}},\ and\ \bibinfo {author}
  {\bibfnamefont {A.}~\bibnamefont {Van~Proeyen}},\ }\bibfield  {title}
  {\bibinfo {title} {{Jordan Frame Supergravity and Inflation in NMSSM}},\
  }\href {https://doi.org/10.1103/PhysRevD.82.045003} {\bibfield  {journal}
  {\bibinfo  {journal} {Phys. Rev. D}\ }\textbf {\bibinfo {volume} {82}},\
  \bibinfo {pages} {045003} (\bibinfo {year} {2010})},\ \Eprint
  {https://arxiv.org/abs/1004.0712} {arXiv:1004.0712 [hep-th]} \BibitemShut
  {NoStop}%
\bibitem [{\citenamefont {Okada}\ \emph {et~al.}(2010)\citenamefont {Okada},
  \citenamefont {Rehman},\ and\ \citenamefont {Shafi}}]{Okada:2010jf}%
  \BibitemOpen
  \bibfield  {author} {\bibinfo {author} {\bibfnamefont {N.}~\bibnamefont
  {Okada}}, \bibinfo {author} {\bibfnamefont {M.~U.}\ \bibnamefont {Rehman}},\
  and\ \bibinfo {author} {\bibfnamefont {Q.}~\bibnamefont {Shafi}},\ }\bibfield
   {title} {\bibinfo {title} {{Tensor to Scalar Ratio in Non-Minimal $\phi^4$
  Inflation}},\ }\href {https://doi.org/10.1103/PhysRevD.82.043502} {\bibfield
  {journal} {\bibinfo  {journal} {Phys.Rev.}\ }\textbf {\bibinfo {volume}
  {D82}},\ \bibinfo {pages} {043502} (\bibinfo {year} {2010})},\ \Eprint
  {https://arxiv.org/abs/1005.5161} {arXiv:1005.5161 [hep-ph]} \BibitemShut
  {NoStop}%
\bibitem [{\citenamefont {Dolgov}\ and\ \citenamefont
  {Kirilova}(1990)}]{Dolgov:1989us}%
  \BibitemOpen
  \bibfield  {author} {\bibinfo {author} {\bibfnamefont {A.~D.}\ \bibnamefont
  {Dolgov}}\ and\ \bibinfo {author} {\bibfnamefont {D.~P.}\ \bibnamefont
  {Kirilova}},\ }\bibfield  {title} {\bibinfo {title} {{On particle creation by
  a time-dependent scalar field}},\ }\href@noop {} {\bibfield  {journal}
  {\bibinfo  {journal} {Sov. J. Nucl. Phys.}\ }\textbf {\bibinfo {volume}
  {51}},\ \bibinfo {pages} {172} (\bibinfo {year} {1990})},\ \bibinfo {note}
  {[Yad. Fiz.51,273(1990)]}\BibitemShut {NoStop}%
\bibitem [{\citenamefont {Traschen}\ and\ \citenamefont
  {Brandenberger}(1990)}]{Traschen:1990sw}%
  \BibitemOpen
  \bibfield  {author} {\bibinfo {author} {\bibfnamefont {J.~H.}\ \bibnamefont
  {Traschen}}\ and\ \bibinfo {author} {\bibfnamefont {R.~H.}\ \bibnamefont
  {Brandenberger}},\ }\bibfield  {title} {\bibinfo {title} {{Particle
  Production During Out-of-equilibrium Phase Transitions}},\ }\href
  {https://doi.org/10.1103/PhysRevD.42.2491} {\bibfield  {journal} {\bibinfo
  {journal} {Phys. Rev.}\ }\textbf {\bibinfo {volume} {D42}},\ \bibinfo {pages}
  {2491} (\bibinfo {year} {1990})}\BibitemShut {NoStop}%
\bibitem [{\citenamefont {Kawai}\ and\ \citenamefont
  {Nakayama}(2016)}]{Kawai:2015lja}%
  \BibitemOpen
  \bibfield  {author} {\bibinfo {author} {\bibfnamefont {S.}~\bibnamefont
  {Kawai}}\ and\ \bibinfo {author} {\bibfnamefont {Y.}~\bibnamefont
  {Nakayama}},\ }\bibfield  {title} {\bibinfo {title} {{Reheating of the
  Universe as holographic thermalization}},\ }\href
  {https://doi.org/10.1016/j.physletb.2016.06.019} {\bibfield  {journal}
  {\bibinfo  {journal} {Phys. Lett.}\ }\textbf {\bibinfo {volume} {B759}},\
  \bibinfo {pages} {546} (\bibinfo {year} {2016})},\ \Eprint
  {https://arxiv.org/abs/1509.04661} {arXiv:1509.04661 [hep-th]} \BibitemShut
  {NoStop}%
\bibitem [{\citenamefont {Kawasaki}\ \emph {et~al.}(2008)\citenamefont
  {Kawasaki}, \citenamefont {Kohri}, \citenamefont {Moroi},\ and\ \citenamefont
  {Yotsuyanagi}}]{Kawasaki:2008qe}%
  \BibitemOpen
  \bibfield  {author} {\bibinfo {author} {\bibfnamefont {M.}~\bibnamefont
  {Kawasaki}}, \bibinfo {author} {\bibfnamefont {K.}~\bibnamefont {Kohri}},
  \bibinfo {author} {\bibfnamefont {T.}~\bibnamefont {Moroi}},\ and\ \bibinfo
  {author} {\bibfnamefont {A.}~\bibnamefont {Yotsuyanagi}},\ }\bibfield
  {title} {\bibinfo {title} {{Big-Bang Nucleosynthesis and Gravitino}},\ }\href
  {https://doi.org/10.1103/PhysRevD.78.065011} {\bibfield  {journal} {\bibinfo
  {journal} {Phys. Rev. D}\ }\textbf {\bibinfo {volume} {78}},\ \bibinfo
  {pages} {065011} (\bibinfo {year} {2008})},\ \Eprint
  {https://arxiv.org/abs/0804.3745} {arXiv:0804.3745 [hep-ph]} \BibitemShut
  {NoStop}%
\bibitem [{\citenamefont {Akrami}\ \emph {et~al.}(2018)\citenamefont {Akrami}
  \emph {et~al.}}]{Akrami:2018odb}%
  \BibitemOpen
  \bibfield  {author} {\bibinfo {author} {\bibfnamefont {Y.}~\bibnamefont
  {Akrami}} \emph {et~al.} (\bibinfo {collaboration} {Planck}),\ }\bibfield
  {title} {\bibinfo {title} {{Planck 2018 results. X. Constraints on
  inflation}},\ }\href@noop {} {\  (\bibinfo {year} {2018})},\ \Eprint
  {https://arxiv.org/abs/1807.06211} {arXiv:1807.06211 [astro-ph.CO]}
  \BibitemShut {NoStop}%
\bibitem [{\citenamefont {Baldini}\ \emph {et~al.}(2018)\citenamefont {Baldini}
  \emph {et~al.}}]{MEGII:2018kmf}%
  \BibitemOpen
  \bibfield  {author} {\bibinfo {author} {\bibfnamefont {A.~M.}\ \bibnamefont
  {Baldini}} \emph {et~al.} (\bibinfo {collaboration} {MEG II}),\ }\bibfield
  {title} {\bibinfo {title} {{The design of the MEG II experiment}},\ }\href
  {https://doi.org/10.1140/epjc/s10052-018-5845-6} {\bibfield  {journal}
  {\bibinfo  {journal} {Eur. Phys. J. C}\ }\textbf {\bibinfo {volume} {78}},\
  \bibinfo {pages} {380} (\bibinfo {year} {2018})},\ \Eprint
  {https://arxiv.org/abs/1801.04688} {arXiv:1801.04688 [physics.ins-det]}
  \BibitemShut {NoStop}%
\bibitem [{\citenamefont {Tumasyan}\ \emph {et~al.}(2021)\citenamefont
  {Tumasyan} \emph {et~al.}}]{CMS:2021kdm}%
  \BibitemOpen
  \bibfield  {author} {\bibinfo {author} {\bibfnamefont {A.}~\bibnamefont
  {Tumasyan}} \emph {et~al.} (\bibinfo {collaboration} {CMS}),\ }\bibfield
  {title} {\bibinfo {title} {{Search for long-lived particles decaying to
  leptons with large impact parameter in proton-proton collisions at $\sqrt{s}$
  = 13 TeV}},\ }\href@noop {} {\  (\bibinfo {year} {2021})},\ \Eprint
  {https://arxiv.org/abs/2110.04809} {arXiv:2110.04809 [hep-ex]} \BibitemShut
  {NoStop}%
\bibitem [{\citenamefont {Felea}\ \emph {et~al.}(2020)\citenamefont {Felea},
  \citenamefont {Mamuzic}, \citenamefont {Mase\l{}ek}, \citenamefont
  {Mavromatos}, \citenamefont {Mitsou}, \citenamefont {Pinfold}, \citenamefont
  {Ruiz~de Austri}, \citenamefont {Sakurai}, \citenamefont {Santra},\ and\
  \citenamefont {Vives}}]{Felea:2020cvf}%
  \BibitemOpen
  \bibfield  {author} {\bibinfo {author} {\bibfnamefont {D.}~\bibnamefont
  {Felea}}, \bibinfo {author} {\bibfnamefont {J.}~\bibnamefont {Mamuzic}},
  \bibinfo {author} {\bibfnamefont {R.}~\bibnamefont {Mase\l{}ek}}, \bibinfo
  {author} {\bibfnamefont {N.~E.}\ \bibnamefont {Mavromatos}}, \bibinfo
  {author} {\bibfnamefont {V.~A.}\ \bibnamefont {Mitsou}}, \bibinfo {author}
  {\bibfnamefont {J.~L.}\ \bibnamefont {Pinfold}}, \bibinfo {author}
  {\bibfnamefont {R.}~\bibnamefont {Ruiz~de Austri}}, \bibinfo {author}
  {\bibfnamefont {K.}~\bibnamefont {Sakurai}}, \bibinfo {author} {\bibfnamefont
  {A.}~\bibnamefont {Santra}},\ and\ \bibinfo {author} {\bibfnamefont
  {O.}~\bibnamefont {Vives}},\ }\bibfield  {title} {\bibinfo {title}
  {{Prospects for discovering supersymmetric long-lived particles with
  MoEDAL}},\ }\href {https://doi.org/10.1140/epjc/s10052-020-7994-7} {\bibfield
   {journal} {\bibinfo  {journal} {Eur. Phys. J. C}\ }\textbf {\bibinfo
  {volume} {80}},\ \bibinfo {pages} {431} (\bibinfo {year} {2020})},\ \Eprint
  {https://arxiv.org/abs/2001.05980} {arXiv:2001.05980 [hep-ph]} \BibitemShut
  {NoStop}%
\bibitem [{\citenamefont {Casas}\ and\ \citenamefont
  {Ibarra}(2001)}]{Casas:2001sr}%
  \BibitemOpen
  \bibfield  {author} {\bibinfo {author} {\bibfnamefont {J.~A.}\ \bibnamefont
  {Casas}}\ and\ \bibinfo {author} {\bibfnamefont {A.}~\bibnamefont {Ibarra}},\
  }\bibfield  {title} {\bibinfo {title} {{Oscillating neutrinos and $\mu \to e,
  \gamma$}},\ }\href {https://doi.org/10.1016/S0550-3213(01)00475-8} {\bibfield
   {journal} {\bibinfo  {journal} {Nucl. Phys. B}\ }\textbf {\bibinfo {volume}
  {618}},\ \bibinfo {pages} {171} (\bibinfo {year} {2001})},\ \Eprint
  {https://arxiv.org/abs/hep-ph/0103065} {arXiv:hep-ph/0103065} \BibitemShut
  {NoStop}%
\bibitem [{\citenamefont {Ibarra}\ and\ \citenamefont
  {Ross}(2004)}]{Ibarra:2003up}%
  \BibitemOpen
  \bibfield  {author} {\bibinfo {author} {\bibfnamefont {A.}~\bibnamefont
  {Ibarra}}\ and\ \bibinfo {author} {\bibfnamefont {G.~G.}\ \bibnamefont
  {Ross}},\ }\bibfield  {title} {\bibinfo {title} {{Neutrino phenomenology: The
  Case of two right-handed neutrinos}},\ }\href
  {https://doi.org/10.1016/j.physletb.2004.04.037} {\bibfield  {journal}
  {\bibinfo  {journal} {Phys.Lett.}\ }\textbf {\bibinfo {volume} {B591}},\
  \bibinfo {pages} {285} (\bibinfo {year} {2004})},\ \Eprint
  {https://arxiv.org/abs/hep-ph/0312138} {arXiv:hep-ph/0312138 [hep-ph]}
  \BibitemShut {NoStop}%
\end{thebibliography}
